\documentclass[superscriptaddress,nobibnotes,amsmath,amssymb,notitlepage,twocolumn,prl]{revtex4-2}

\usepackage{pdfpages}
\usepackage{etoolbox} 

\usepackage{mathptmx}
\usepackage{graphicx}
\usepackage{multirow}
\usepackage{hyperref}
\hypersetup{colorlinks=true, linkcolor=blue, citecolor=blue, urlcolor=blue}

\graphicspath{{./img/}}

\newcommand{\beq}[1]{\begin{equation}\label{#1}}
\newcommand{\eep}{\;.\end{equation}}
\newcommand{\eec}{\;,\end{equation}}
\newcommand{\eeq}{\end{equation}}

\newcommand*{\headingSectionPRL}[1]{\belowpdfbookmark{#1}{#1}{\textit{#1.---}}\ignorespaces}
\let\section\headingSectionPRL

\newcommand*\chem[1]{\ensuremath{\mathrm{#1}}}

\renewcommand{\a}{\alpha}

\DeclareMathAlphabet{\mathcal}{OMS}{cmsy}{m}{n} 
\newcommand{\C}{\mathcal{C}} 
\newcommand{\Ef}{\mathcal{E}}   
\newcommand{\V}{\mathcal{V}}  

\renewcommand{\vec}[1]{{\bf #1}}
\newcommand{\x}{\vec{x}}
\renewcommand{\P}{\vec{P}}

\newcommand{\Vstack}{\V_{\rm stack}}

\newcommand{\Bc}{B_{\rm c}}

\begin{document}

\makeatletter
\patchcmd{\@outputpage@head}{\@ifx{\LS@rot\@undefined}{}{\LS@rot}}{}{}{}
\makeatother

\title{Stacking-engineered ferroelectricty and multiferroic order in van der Waals magnets}


\newcommand{\HarvardPhysics}{Department of Physics, Harvard University, Cambridge, Massachusetts 02138, USA}
\newcommand{\HarvardSeas}{John A.~Paulson School of Engineering and Applied Sciences, Harvard University, Cambridge, Massachusetts 02138, USA}
\newcommand{\Oviedo}{Departamento de Física, Universidad de Oviedo, 33007 Oviedo, Spain}
\newcommand{\CINN}{Centro de Investigación en Nanomateriales y Nanotecnología, Universidad de Oviedo-CSIC, 33940 El Entrego, Spain}
\newcommand{\Liege}{Theoretical Materials Physics, Q-MAT, University of Liège, B-4000 Sart Tilman, Belgium}
\newcommand{\MIT}{Department of Physics, Massachusetts Institute of Technology, Cambridge, MA, USA}

\author{Daniel Bennett}
\email{dbennett@seas.harvard.edu}
\affiliation{\HarvardSeas}

\author{Gabriel Martínez-Carracedo}
\affiliation{\Oviedo}
\affiliation{\CINN}

\author{Xu He}
\affiliation{\Liege}

\author{Jaime Ferrer}
\affiliation{\Oviedo}
\affiliation{\CINN}

\author{Philippe Ghosez}
\affiliation{\Liege}

\author{Riccardo Comin}
\affiliation{\MIT}

\author{Efthimios Kaxiras}
\affiliation{\HarvardSeas}
\affiliation{\HarvardPhysics}

\begin{abstract}
Two-dimensional (2D) materials that exhibit spontaneous magnetization, polarization or strain (referred to as ferroics) have the potential to revolutionize nanotechnology by enhancing the multifunctionality of nanoscale devices.
However, multiferroic order is difficult to achieve, requiring complicated coupling between electron and spin degrees of freedom.
We propose a universal method to engineer multiferroics from van der Waals magnets by taking advantage of the fact that changing the stacking between 2D layers can break inversion symmetry, resulting in ferroelectricity as well as magnetoelectric coupling. 
We illustrate this concept using first-principles calculations in bilayer \chem{NiI_2}, which can be made ferroelectric upon rotating two adjacent layers by $180^{\circ}$ with respect to the bulk stacking. 
Furthermore, we discover a novel strong magnetoelectric coupling between the interlayer spin order and interfacial electronic polarization.
Our approach is not only general but also systematic, and can enable the discovery of a wide variety of 2D multiferroics with strong magnetoelectric coupling.

\end{abstract}

\maketitle


\section{Introduction}
The ability to synthesize stable layered van der Waals (vdW) materials has opened up new possibilities for mechanically-assembled stacks of heterostructures and led to significant advances in materials science \cite{novoselov20162d,geim2013van,gibertini2019magnetic}.
These new artificial materials display a range of unprecedented multifunctional properties \cite{khan2020recent,glavin2020emerging,kumbhakar2023prospective}, one notable example being ferroic orders such as ferromagnetism \cite{huang2017layer,gong2017discovery} and ferrroelectricity \cite{li2017binary,stern2020interfacial,yasuda2021stacking}.
2D magnetic materials show great promise as a new class of magnets because their properties can be controlled through the application of external stimuli beyond temperature and magnetic field, such as mechanical strain or gate voltage \cite{mermin1966absence,vsivskins2022nanomechanical,lv2019strain,martinez2023electrically}.
Recently a new type of ferroelectricity in vdW materials was proposed \cite{li2017binary} and experimentally observed \cite{stern2020interfacial,yasuda2021stacking,wang2022interfacial,weston2022interfacial,ko2023operando,molino2023ferroelectric,van2024engineering}.
For vdW materials such as hexagonal boron nitride (hBN) or transition metal dichalcogenides (TMDs) where centrosymmetry is broken by stacking-engineering, an electronic out-of-plane polarization $P_{\perp}$ occurs through interlayer charge transfer. 
The magnitude and orientation of $P_{\perp}$ is determined by the relative stacking between the layers and can be switched by a relative sliding, resulting in `vdW ferroelectricity' \cite{yasuda2021stacking,yasuda2024ultrafast,bian2024developing}. 

The functional properties of vdW materials can be further tuned by introducing a relative twist or lattice mismatch between neighboring layers, resulting in a moir\'e superlattice on a scale much larger than the underlying crystal periodicity.
This has been the source of many interesting and unconventional emergent phenomena, one notable example being superconductivity in `magic angle' twisted bilayer graphene \cite{Bistritzer2011,cao2018unconventional,cao2018correlated,bennett2024twisted}.
Creating a moir\'e superlattice from a vdW ferroelectric results in the formation of moir\'e polar domains (MPDs) \cite{bennett2022electrically,bennett2022theory}.
These MPDs have been experimentally shown to result in ferroelectricity \cite{yasuda2021stacking,ko2023operando} through the asymmetric change in domain size in response to an applied electric field.
The MPDs also have an in-plane polarization component $\P_{\parallel}$ and as a result they have nontrivial topology \cite{bennett2023polar,bennett2023theory,jankowski2024polarization,vu2024imaging}, analogous to the polar textures observed in oxide perovskites \cite{das2019observation,han2022high,junquera2023topologicaly,sanchez20242d}.
Similar topological magnetic textures have been proposed and reported by twisting vdW magnets, such as \chem{CrI_3} \cite{song2021direct,fumega2023moire}.

The coexistence or interplay between ferroelectricity and magnetism can give rise to multiferroic order \cite{spaldin2005renaissance,spaldin2019advances}, which is promising for enhanced functionality in nanodevices.
While 2D multiferroics are less common than bulk counterparts such as \chem{BiFeO_3} \cite{wang2003epitaxial}, the vdW magnet \chem{NiI_2} \cite{liu2020vapor,ju2021possible,Santos2022} has shown signals of multiferroic order down to the monolayer limit \cite{song2022evidence}.
The proposed mechanism in \chem{NiI_2} is the appearance of improper ferroelectricity in response to long-range magnetic order which breaks inversion symmetry \cite{mostovoy2006ferroelectricity,xiang2011general,hohenberg2015introduction,fumega2022microscopic,fumega2023moire}.
There are a few disadvantages to this relatively complicated magnetoelectric coupling mechanism.
The strength of the coupling depends on the strength of the Rashba spin-orbit coupling (SOC) in a material and may be weak in materials that do not contain heavy elements. 
Moreover, it also requires magnetization gradients such as complex spin textures.

\begin{figure*}[t]
\centering
\includegraphics[width=\linewidth]{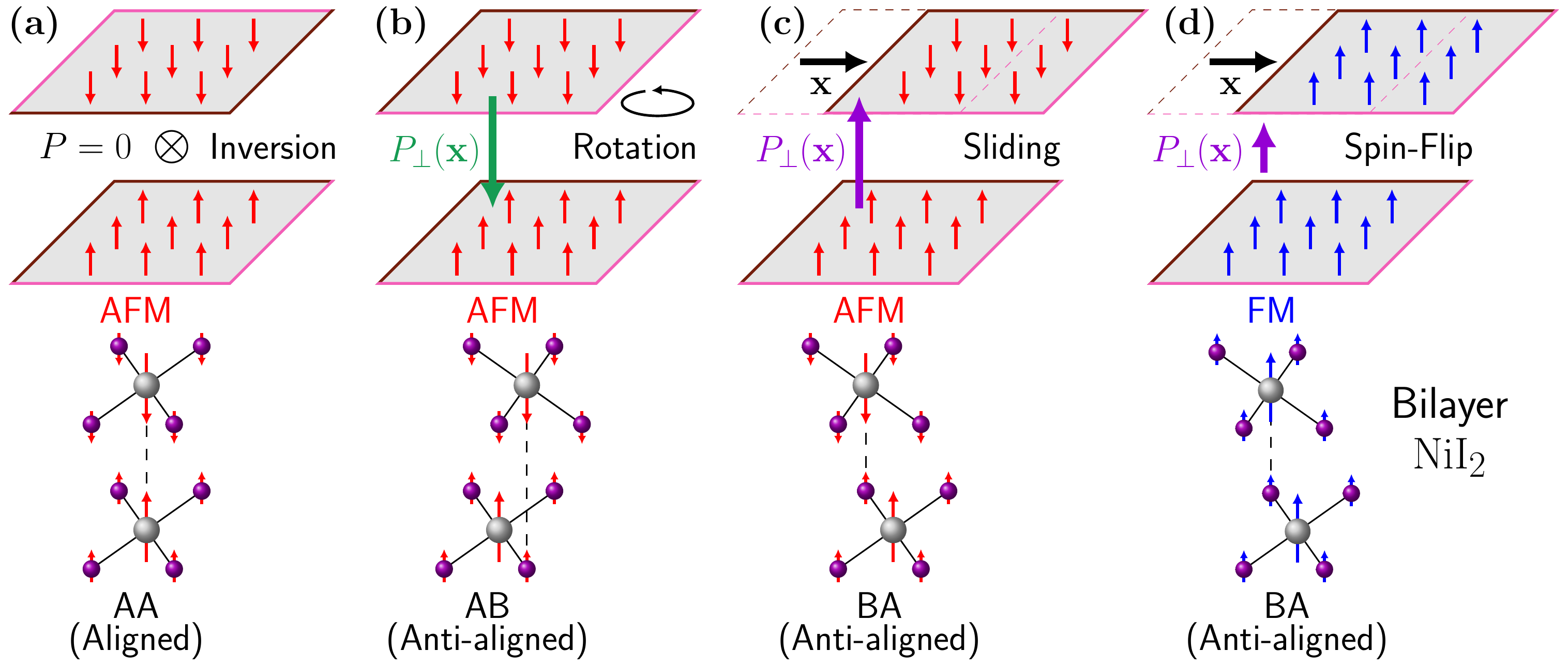}
\caption{
Illustration of tuning multiferroic order and magnetoelectric coupling in vdW magnets with stacking engineering.
{\bf (a)} A vdW bilayer with an AFM interlayer spin configuration is sketched (spins denoted by red arrows),
with color-coded edges to help identify the relative orientation of the layers.
The bilayer has an inversion center and is nonpolar.
{\bf (b)} Rotating the layers by $180^{\circ}$ breaks the inversion center, resulting in a stacking dependent polarization $P_{\perp}(\x)$, denoted by the green (down) and purple (up) arrows.
{\bf (c)} A relative sliding $\x$ between the layers inverts the polarization.
{\bf (d)} Changing the interlayer spin order to FM (spins denoted by blue arrows) affects the magnitude of the polarization, but not the direction.
The corresponding stackings in bilayer \chem{NiI_2} are illustrated below in each case.
There is a relative shift between the layers along the unit cell diagonal: $\x = x ({\bf a}_1 + {\bf a}_2)$, where $x_{\rm AA} = 0$, $x_{\rm AB} = \frac{1}{3}$ and $x_{\rm BA} = \frac{2}{3}$.
Most of the magnetic moment is concentrated on the Ni atoms, with a small magnetic moment transferred to the I atoms.
}
\label{Fig1}
\end{figure*}

While the coexistence of ferroic orders has previously been predicted in other vdW magnets \cite{li2017binary,sivadas2018stacking,ji2023general,xun2024coexisting}, significant coupling between polarization and magnetization, which would enhance multifunctionality \cite{spaldin2005renaissance}, is not guaranteed.
Here we discover a novel strong magnetoelectric coupling in artificially stacked vdW magnets, between the interlayer electronic polarization and spin ordering.
In the model system we use to illustrate this effect, NiI$_2$, we find that changing the interlayer spin order modifies the electronic structure and can change the electronic polarization by a factor of two; this in turn leads to additional jumps in the ferroelectric hysteresis loop, implying that ferroelectric polarization may be tuned using a magnetic field, and magnetization may be tuned using an electric field.

We propose a simple and general approach for engineering multiferroics with magnetoelectric coupling from vdW magnets through artificial stacking (see Fig.~\ref{Fig1}). 
This approach has already been successfully been applied to fabricate vdW ferroelectrics \cite{yasuda2021stacking,wang2022interfacial,ko2023operando} mechanically using the tear and stack method. 
Using bilayer \chem{NiI_2} as an example, we show with {\it ab initio} calculations that by rotating one of the layers by 180$^{\circ}$ with respect to the natural bulk stacking, the bilayer becomes ferroelectric (see Fig.~\ref{Fig2}).
Whether or not a system can be made ferroelectric can be determined simply from knowledge of the space groups of the different stackings \cite{ji2023general}.
It turns out that the symmetries of bilayers of \chem{NiI_2} and of TMDs in the H1 structure (such as \chem{MoS_2}) are {\it exactly} the same upon rotation by 180$^{\circ}$ with respect to the natural bulk stacking (see Table \ref{table:space-group}).
This feature makes their ferroelectric properties identical.

\begin{table}[t]
\renewcommand*{\arraystretch}{2}
\setlength{\tabcolsep}{3pt}
\begin{center}
\begin{tabular}{| c | c | c | c | c |}
\hline\hline
\multirow{2}{*}{Stacking} & \multicolumn{2}{c | }{\chem{NiI_2}} & \multicolumn{2}{c |}{\chem{MoS_2}} \\ \cline{2-5}
 & {\bf Aligned} & Anti-aligned & Aligned & {\bf Anti-aligned} \\ \hline
 AA & $P\bar{3}m1$ (\#164) & $P\bar{6}m2$ (\#187) & $P\bar{6}m2$ (\#187) & $P\bar{3}m1$ (\#164) \\ \hline
 AB & $P\bar{3}m1$ (\#164) & $P3m1$ (\#156) & $P3m1$ (\#156) & $P\bar{3}m1$ (\#164) \\ \hline
 DW & $C2/m$ (\#12) & $Abm2$ (\#39) & $Abm2$ (\#39) & $C2/m$ (\#12) \\ \hline
 BA & $P\bar{3}m1$ (\#164) & $P3m1$ (\#156) & $P3m1$ (\#156) & $P\bar{3}m1$ (\#164) \\ \hline
\hline
\end{tabular}
\end{center}
\caption{
Space group analysis of the high-symmetry relative stackings in bilayer \chem{NiI2} and \chem{MoS_2}. The bold font indicates the natural bulk stacking.}
\label{table:space-group}
\end{table}


\section{Results}
We performed first-principles density functional theory (DFT) calculations to simulate the properties of bilayer \chem{NiI_2} \cite{SM}. 
We considered the aligned stacking which occurs naturally in bulk \chem{NiI_2}, as well as the artificial anti-aligned stacking (see Fig.~\ref{Fig1}),
with antiferromagnetic (AFM) and ferromagnetic (FM) interlayer spin configurations.
We emphasize that throughout this work, AFM and FM always refer to the {\it interlayer} alignment of spins and not the alignment within a single layer.
Because the wavelength of the spin spiral in \chem{NiI2} is relatively long (around 7 unit cells \cite{fumega2022microscopic}), we consider uniform arrangements of spins pointing in the out-of-plane direction in each layer in order to avoid calculations involving large supercells.

\begin{figure}[t]
\centering
\includegraphics[width=\linewidth]{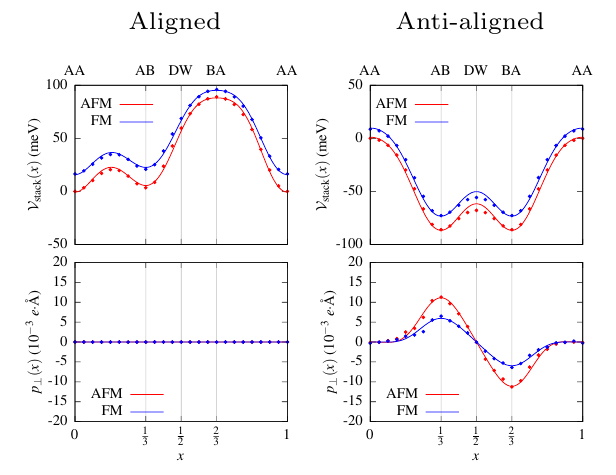}
\caption{
Stacking energy $\Vstack$ and out-of-plane dipole moment $p_{\perp}$ for aligned and anti-aligned bilayer \chem{NiI2}, 
as a function of relative stacking $x$ along the unit cell diagonal.
The labels AA, AB/BA and DW refer to the relative stackings, with AA (aligned) and AB/BA (ani-aligned) sketched in Fig.~\ref{Fig1}.
Results for an AFM (FM) interlayer spin order are shown in red (blue). 
The dots show results from first-principles calculations and the solid lines show fits of the data to $\C_3$-symmetric basis functions, such as Eq.~\eqref{eq:fit-Pz} for $p_{\perp}$.
}
\label{Fig2}
\end{figure}

Fig.~\ref{Fig2} shows the stacking energy $\Vstack$ and out-of-plane electronic dipole moment $p_{\perp}$ as a function of relative displacement $x$ between the layers along the unit cell diagonal: $\x = x ({\bf a}_1 + {\bf a}_2)$ where ${\bf a}_i$ are the lattice vectors. 
Results for aligned and anti-aligned stackings as well as AFM (red) and FM (blue) interlayer spin configurations are shown \cite{SM}.
The unit cell diagonal contains all of the high symmetry stacking configurations, namely AA ($x=0,1$), AB ($x = \tfrac{1}{3}$) and BA ($x = \tfrac{2}{3}$), and the domain wall (DW) stacking ($x = \tfrac{1}{2}$).
The AFM configuration was found to be lower in energy for all stackings in both aligned and anti-aligned \chem{NiI_2}.
The aligned bilayer is non-polar for all stacking configurations as the inversion center between the layers is preserved upon sliding.
For the anti-aligned stacking, the bilayer attains a stacking-dependent electronic dipole moment which is largest and of opposite sign for the energetically favorable AB and BA stackings, and is comparable to the dipole moments of other vdW ferroelectrics \cite{SM} which, while weak compared to conventional ferroelectrics such as oxide perovskites, can still be measured, switched, and can be used to dope proximal materials \cite{stern2020interfacial,yasuda2021stacking,wang2022interfacial,weston2022interfacial,ko2023operando,molino2023ferroelectric,van2024engineering,yasuda2024ultrafast,bian2024developing}.
The polarization being an odd function of stacking and respecting the $\C_3$ rotation symmetry of the crystal takes the form \cite{bennett2022theory}:
\beq{eq:fit-Pz}
p_{\perp}(x,y) = p^{\text{odd}}_1 \bigg[  \sin{(2\pi x)} + \sin{(2\pi y)} - \sin{(2\pi (x+y))} \bigg]
\eec
with $\x = (x,y)$ in fractional coordinates of the unit cell lattice constant.
$p^{\text{odd}}_1$ is a coefficient to be fit to DFT calculations.
Additionally, anti-aligned \chem{NiI_2} also has a stacking-dependent {\it in-plane} polarization similar to rhombohedral hBN and TMDs \cite{SM}.

\begin{figure}[t]
\centering
\includegraphics[width=\linewidth]{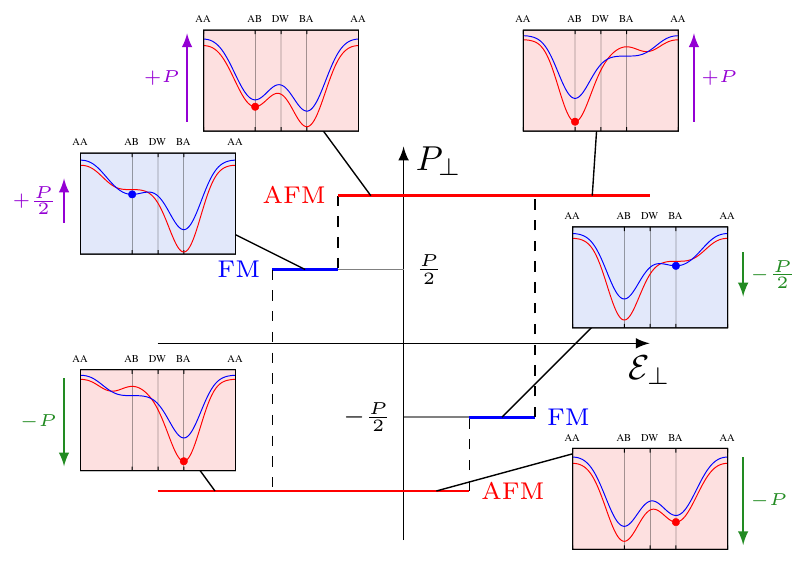}
\caption{
Illustration of the ferroelectric hysteresis loop for anti-aligned bilayer \chem{NiI2}. 
The free energies of the AFM (red) and FM (blue) interlayer spin configurations are sketched as a function of stacking, at six different points around the hysteresis loop.
The dot shows the state that \chem{NiI_2} adopts for each electric field value, which is possibly metastable.
The red (blue) background indicates that \chem{NiI_2} has an AFM (FM) interlayer spin order.
The purple (green) arrows indicate that \chem{NiI_2} has polarization up (down).
The jumps that decrease the magnitude of the polarization but do not change its sign are driven by a change in interlayer spin order (AFM $\rightarrow$ FM).
The jumps which change the sign in polarization are driven by interlayer sliding (AB $\rightarrow$ BA) and are also accompanied by a change in interlayer spin order (FM $\rightarrow$ AFM).
}
\label{Fig3}
\end{figure}

Fig.~\ref{Fig2} shows that the magnitude of the polarization in anti-aligned \chem{NiI_2} is sensitive to the interlayer spin configuration.
Upon switching from AFM to FM the polarization decreases in magnitude by approximately a factor of 2.
Fig.~\ref{Fig3} illustrates the ferroelectric hysteresis loop in anti-aligned \chem{NiI_2} including the effects of this magnetoelectric coupling.
In addition to the first-order interlayer sliding jumps in polarization there are also jumps in the magnitude of polarization, but not in sign, driven by a change in interlayer spin order.
Thus, the spin order in anti-aligned \chem{NiI_2} can be switched using an applied electric field and the magnitude of the polarization can be tuned using a magnetic field.
This can be explained by considering the free energy:
\beq{eq:V-alpha}
\V^{\a}(\x,\Ef) = \Vstack^{\a}(\x) - \Ef\cdot p^{\a}(\x)
\eec
where $\a =$ AFM, FM,
which is sketched in Fig.~\ref{Fig3} at different points around the hysteresis loop.
Applying an electric field $\Ef$ which is anti-aligned with the dipole moment (AB/BA), the FM configuration eventually becomes lower in energy.
This results in a change in the interlayer spin order driven by a decrease in polarization.
Increasing the electric field further, the other stacking configuration will eventually become preferable (BA/AB), resulting in polarization reversal through interlayer sliding.

The intralayer magnetic exchange parameters were obtained  using the relativistic LKAG formalism \cite{liechtenstein1987local} within a non-orthogonal localized basis framework, implemented in the {\sc grogu} code \cite{grogu,SM}. 
The exchange and intra-atomic anisotropy tensors to any desired neighbor shell can be calculated, corresponding to the following Heisenberg Hamiltonian:  
\beq{eq:H-intra}
H_{\rm intra}=\,\frac{1}{2}\,\sum_{i\neq j}\,J^{ij}_{\parallel}\,{\bf e}_i\cdot{\bf e}_j + 
\frac{1}{2}\,\sum_{i\neq j}\,{\bf e}_i\,J_{S,\parallel}^{ij}\,{\bf e}_j+
\,\frac{1}{2}\,\sum_{i\neq j}\,{\bf D}^{ij}\cdot({\bf e}_i\times{\bf e}_j)
\eep
Here ${\bf e}_i$ is a unit vector pointing along the spin angular momentum of the localized magnetic entity placed at site $i$. 
$J^{ij}_{\parallel}$, $J_{S,\parallel}^{ij}$ and ${\bf D}^{ij}$ are the isotropic exchange constants, symmetric traceless exchange tensor and Dzyaloshinskii-Moriya (DM) vectors between sites $i$ and $j$, respectively. 
The interlayer isotropic exchange parameter $J_{\perp}$ was obtained from total energy differences between the FM and AFM bilayer spin alignments, mapped onto the effective Hamiltonian:
\beq{eq:H-inter}
H_{\rm inter }=J_{\perp} {\bf e}_1 \cdot {\bf e}_2 
\eec
where ${\bf e}_{1}$ and ${\bf e}_{2}$ are unit vectors describing the spin directions in the first and second layers of bilayer \chem{NiI_2}.
The interlayer exchange changes by as much as a factor of 2 as a function of stacking \cite{SM} which suggests that the magnetic transition temperature is also stacking-dependent.
The interlayer spin order does not change upon changing the stacking in contrast to other vdW magnets such as \chem{CrI_3} \cite{sivadas2018stacking,fumega2023moire,sun2023theoretical} and \chem{GdI_2} \cite{xun2024coexisting}: the AFM spin order is always energetically preferable at zero electric field.

\begin{figure}[t]
\centering
\includegraphics[width=\linewidth]{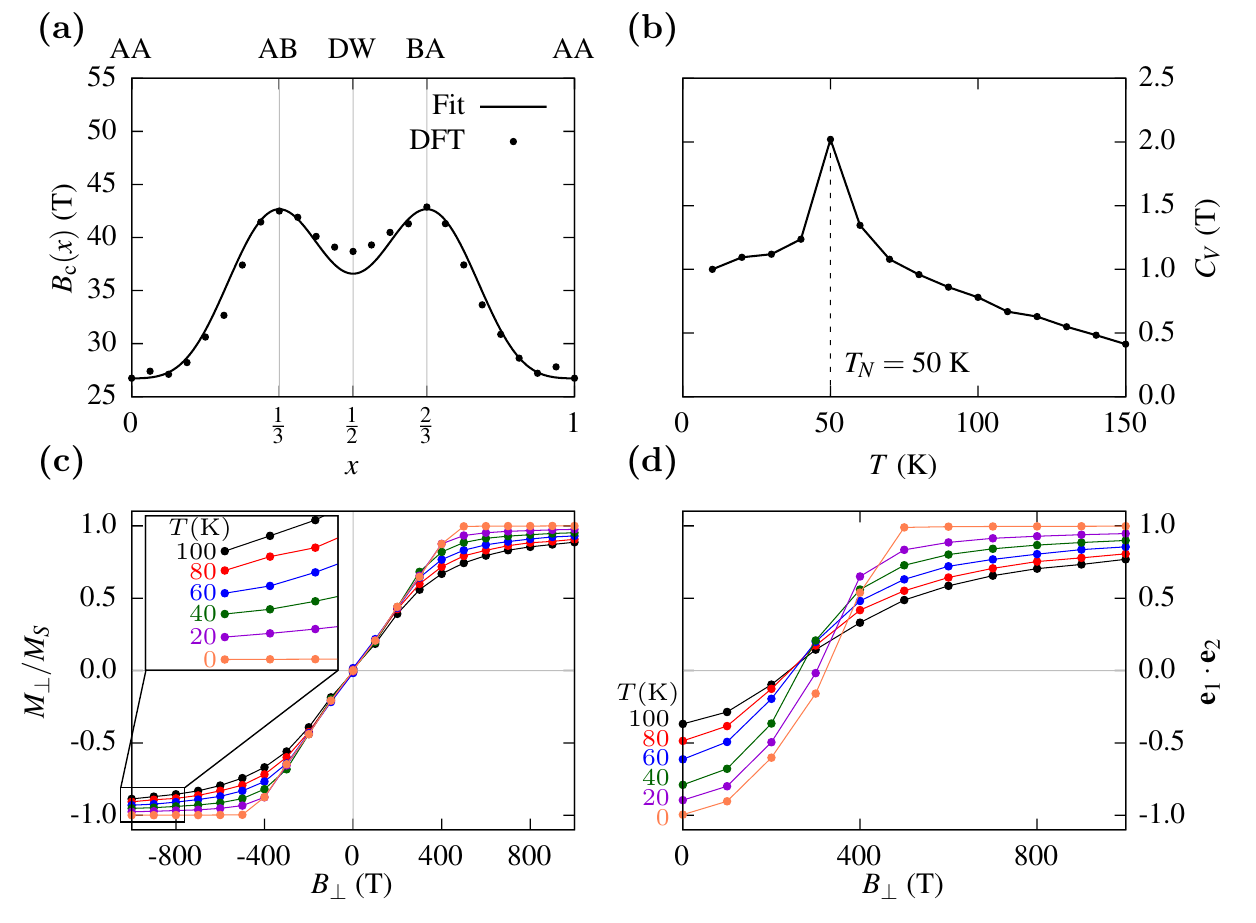}
\caption{
{\bf (a)} Estimate of the coercive magnetic field $\Bc(x)$ in anti-aligned bilayer \chem{NiI_2}. 
The points show the values of $\Bc$ obtained from DFT calculations at each stacking $x$.
The solid line shows the values obtained from the fits to the stacking energies in Fig.~\ref{Fig2}.
{\bf (b)} Specific heat $C_V$ of the system as a function of temperature $T$ from MC simulations.
$C_V$ is normalized such that $C_V(T\to 0) = 1$.
The peak at 50 K indicates the N{\'e}el temperature $T_N$.
{\bf (c)} Average out-of-plane magnetization $M_{\perp}$, in units of the spontaneous magnetization $M_{\rm S}$ as a function of perpendicular magnetic field $B_{\perp}$ and temperature $T$.
{\bf (d)} Average interlayer spin order ${\bf e}_1 \cdot {\bf e}_2$ as a function of perpendicular magnetic field $B_{\perp}$ and temperature $T$.
}
\label{Fig4}
\end{figure}

By solving ${\V^{\rm AFM}(\x_{\rm AB},\Ef_{\rm c}) = \V^{\rm FM}(\x_{\rm AB},\Ef_{\rm c})}$ for $\Ef_{\rm c}$, the electric field required to change the interlayer spin order is $\Ef_{\rm c} = 2.77$ V/\AA~for the AB/BA stackings in anti-aligned bilayer \chem{NiI_2}.
Similarly, the coercive magnetic field $\Bc$ required to switch from AFM to FM spin-order in anti-aligned \chem{NiI_2} and tune the electronic polarization is shown in Fig.~\ref{Fig4} (a). 
$\Bc$ modulates with stacking, ranging between 25--40 T.
In order to investigate the switching dynamics in more detail, we perform classical Monte Carlo (MC) calculations with a Heisenberg Hamiltonian to simulate anti-aligned bilayer \chem{NiI_2} in the AB stacking using a $64\times 64$ supercell \cite{SM}.
The intralayer magnetic interaction is frustrated due to the competition between the FM nearest neighbor (NN) exchange interactions and the AFM 3$^{\rm rd}$ NN exchange interaction. 
The ground state from MC simulations is a spin spiral state with the in-plane spin and AFM interlayer ordering, as was found in Ref.~\cite{amoroso2020spontaneous}.
By measuring the specific heat as a function of temperature, we find a N{\'e}el temperature of $T_N = 50$ K (Fig.~\ref{Fig4} (b)).
The average out-of-plane magnetization $M_{\perp}$ in response to an out-of-plane magnetic field $B_{\perp}$ is shown in Fig.~\ref{Fig4} (c), for temperatures ranging from 0 K to 100 K.
The magnetic field required to align the magnetic moment is $B_{\perp} \sim 800$ T at $T=0$ K, and is reduced to about 200 T at finite temperature.
In order to consider the magnetic field required to change the interlayer spin order, we calculate ${\bf e}_1 \cdot {\bf e}_2$, averaged over the supercell, as a function of magnetic field and temperature (Fig.~\ref{Fig4} (d)): the interlayer spin order changes when ${\bf e}_1 \cdot {\bf e}_2$ changes sign.
The coercive magnetic field required to change the interlayer spin order is also sensitive to temperature.
Comparing Figs.~\ref{Fig4} (c) and (d), the coercive field required to change the interlayer spin order is approximately half the field required to align all of the spins with the field (destroy the spin spiral state).


\section{Discussion and conclusions}
We propose that vdW magnets can be made ferroelectric by fabricating multilayers in artificial stackings which break inversion symmetry, and we illustrate this concept using bilayer \chem{NiI_2} as an example.
This approach can be generalized to other vdW magnets, although the resulting polar properties may differ depending on the crystal symmetries \cite{poudel2023creating}.
In addition, stacking-engineering also results in a novel strong magnetoelectric coupling in vdW magnets without centrosymmetry.
The magnetoelectric coupling occurs between interfacial polarization arising from interlayer charge transfer and the interlayer spin order in the bilayer: changes in the electronic structure induced by changes in the interlayer spin ordering modify the electronic polarization.
As a case in point, we find a new type of multiferroic order in bilayer \chem{NiI_2} upon rotation of the layers to the anti-aligned stacking. 
We predict additional jumps in the ferroelectric hysteresis loop, with the polarization changing in magnitude by a factor of two as a result of changes in the interlayer spin order.
While we expect the magnetoelectric coupling to be general to vdW magnets, determined only by symmetry, the existence of the additional jumps in the hysteresis loop (Fig.~\ref{Fig3}) is determined by the relative sizes of the energy barriers separating the polar stackings, and the AFM/FM interlayer spin orders, which may be material dependent.
Additionally, this balance may be sensitive to the switching mechanisms, as discussed below.

The coercive fields required to achieve this switching are predicted to be unrealistically large by our
calculations, with electric fields of order 1 V/\AA~and magnetic fields of order 10--100 T (similar values were predicted in Ref.~\cite{amoroso2020spontaneous}).
We note that the large coercive fields obtained in DFT/MC calculations correspond to the physical mechanism of
polarization/magnetization reversal. Experimentally however, switching occurs due to nano-domain nucleation 
and subsequent growth due to domain wall propagation.
This switching mechanism needs to overcome much lower energy barriers and therefore requires much smaller coercive 
fields, typically below $\Ef = 0.2$ V/nm for vdW ferroelectrics \cite{yasuda2021stacking,wang2022interfacial,ko2023operando} and $B = 1$ T for vdW magnets \cite{wen2020tunable,huang2017layer,fei2018two}.

We expect that our predictions of coupled multiferroic order (Fig.~\ref{Fig3}) may be validated using similar experimental techniques used to characterize vdW ferroelectrics and magnets.
Techniques to measure ferroelectric hysteresis in rhombohedral hBN \cite{yasuda2021stacking,yasuda2024ultrafast} and TMDs \cite{bian2024developing} may be used to detect additional jumps in polarization induced by spin flips.
The polarization in \chem{NiI_2} can be probed through bulk photovoltaic effect (BPVE) measurements, which can also provide information on the role of domain walls in switching \cite{song2022evidence}.
Additionally, polar domains can be imaged using piezoresponse force microscopy (PFM) measurements \cite{stern2020interfacial}, which can be done under a magnetic field \cite{keeney2013magnetic}, or transmission electron microscopy (TEM) measurements \cite{ko2023operando}.
For ultrathin multiferroics, magneto-optical measurements such as circular dichroism (CD) and magneto-optical Kerr effect (MOKE) can be used to determine the magnetization as well as probe magnetic textures \cite{song2022evidence}.

In the present work we considered the interplay between electronic polarization and interlayer spin order assuming homogeneous spin textures within each layer.
The coupling between polarization and complex spin textures may lead to more interesting and exotic multiferroic coupling. 
For example, suppose the spin spirals in both layers are offset from one another, which can be described by a relative phase shift.
If the relative phase shift is the 0 ($\pi$), the interlayer spin order is FM (AFM).
By using an electric field to switch the polarization through interlayer sliding, the relative phase between the two spin spirals and hence the interlayer spin order may also change.

Finally, we propose that stacking vdW magnets to form moir\'e superlattices may be a route to engineering multiferroic topological structures, in which polar and spin topologies are coupled and may both be controlled with magnetic and electric fields.
Considering the out-of-plane and in-plane polarization in anti-aligned bilayer \chem{NiI_2}, a twist by a small angle about the $180^{\circ}$ orientation will result in a regular network of polar merons and antimerons (winding numbers $Q\pm\frac{1}{2}$ in each domain) \cite{bennett2023polar,SM}.
Additionally, it is known that magnetic skyrmions can spontaneously form in \chem{NiI_2} in the few-layer limit \cite{amoroso2020spontaneous}.
By precisely engineering a moir\'e superlattice of period commensurate with the magnetic skyrmion wavelength, one may achieve the coexistence and coupling of polar and magnetic topological structures in a single device.

\section{Acknowledgements}
D.B.~and E.K.~acknowledge the US Army Research Office (ARO) MURI project under grant No.~W911NF-21-0147 and the Simons Foundation award No.~896626.
G.M.-C.~and J.F.~have been funded by MCIN/AEI/10.13039/501100011033/FEDER, UE via project PID2022-137078NB-100 and by Asturias FICYT under grant AYUD/2021/51185 with the support of FEDER funds. 
G.M.-C.~has been supported by Programa ``Severo Ochoa'' de Ayudas para la investigación y docencia del Principado de Asturias.
X.H.~and Ph.G.~acknowledge financial support from F.R.S.-FNRS Belgium through the PDR project PROMOSPAN (Grant No.~T.0107.20).
R.C.~acknowledges support by the Department of Energy, Office of Science, Office of Basic Energy Sciences, under Award No.~DE-SC0019126.
J.F.~acknowledges a discussion with A.~Hierro on the role of magnetization reversal versus domain growth in experimental magnetic materials.


\begin{thebibliography}{72}%
\makeatletter
\providecommand \@ifxundefined [1]{%
 \@ifx{#1\undefined}
}%
\providecommand \@ifnum [1]{%
 \ifnum #1\expandafter \@firstoftwo
 \else \expandafter \@secondoftwo
 \fi
}%
\providecommand \@ifx [1]{%
 \ifx #1\expandafter \@firstoftwo
 \else \expandafter \@secondoftwo
 \fi
}%
\providecommand \natexlab [1]{#1}%
\providecommand \enquote  [1]{``#1''}%
\providecommand \bibnamefont  [1]{#1}%
\providecommand \bibfnamefont [1]{#1}%
\providecommand \citenamefont [1]{#1}%
\providecommand \href@noop [0]{\@secondoftwo}%
\providecommand \href [0]{\begingroup \@sanitize@url \@href}%
\providecommand \@href[1]{\@@startlink{#1}\@@href}%
\providecommand \@@href[1]{\endgroup#1\@@endlink}%
\providecommand \@sanitize@url [0]{\catcode `\\12\catcode `\$12\catcode
  `\&12\catcode `\#12\catcode `\^12\catcode `\_12\catcode `\%12\relax}%
\providecommand \@@startlink[1]{}%
\providecommand \@@endlink[0]{}%
\providecommand \url  [0]{\begingroup\@sanitize@url \@url }%
\providecommand \@url [1]{\endgroup\@href {#1}{\urlprefix }}%
\providecommand \urlprefix  [0]{URL }%
\providecommand \Eprint [0]{\href }%
\providecommand \doibase [0]{http://dx.doi.org/}%
\providecommand \selectlanguage [0]{\@gobble}%
\providecommand \bibinfo  [0]{\@secondoftwo}%
\providecommand \bibfield  [0]{\@secondoftwo}%
\providecommand \translation [1]{[#1]}%
\providecommand \BibitemOpen [0]{}%
\providecommand \bibitemStop [0]{}%
\providecommand \bibitemNoStop [0]{.\EOS\space}%
\providecommand \EOS [0]{\spacefactor3000\relax}%
\providecommand \BibitemShut  [1]{\csname bibitem#1\endcsname}%
\let\auto@bib@innerbib\@empty
\bibitem [{\citenamefont {Novoselov}\ \emph {et~al.}(2016)\citenamefont
  {Novoselov}, \citenamefont {Mishchenko}, \citenamefont {Carvalho},\ and\
  \citenamefont {Castro~Neto}}]{novoselov20162d}%
  \BibitemOpen
  \bibfield  {author} {\bibinfo {author} {\bibfnamefont {K.~S.}\ \bibnamefont
  {Novoselov}}, \bibinfo {author} {\bibfnamefont {A.}~\bibnamefont
  {Mishchenko}}, \bibinfo {author} {\bibfnamefont {A.}~\bibnamefont
  {Carvalho}}, \ and\ \bibinfo {author} {\bibfnamefont {A.}~\bibnamefont
  {Castro~Neto}},\ }\href {\doibase https://doi.org/10.1126/science.aac9439}
  {\bibfield  {journal} {\bibinfo  {journal} {Science}\ }\textbf {\bibinfo
  {volume} {353}},\ \bibinfo {pages} {aac9439} (\bibinfo {year}
  {2016})}\BibitemShut {NoStop}%
\bibitem [{\citenamefont {Geim}\ and\ \citenamefont
  {Grigorieva}(2013)}]{geim2013van}%
  \BibitemOpen
  \bibfield  {author} {\bibinfo {author} {\bibfnamefont {A.~K.}\ \bibnamefont
  {Geim}}\ and\ \bibinfo {author} {\bibfnamefont {I.~V.}\ \bibnamefont
  {Grigorieva}},\ }\href {\doibase https://doi.org/10.1038/nature12385}
  {\bibfield  {journal} {\bibinfo  {journal} {Nature}\ }\textbf {\bibinfo
  {volume} {499}},\ \bibinfo {pages} {419} (\bibinfo {year}
  {2013})}\BibitemShut {NoStop}%
\bibitem [{\citenamefont {Gibertini}\ \emph {et~al.}(2019)\citenamefont
  {Gibertini}, \citenamefont {Koperski}, \citenamefont {Morpurgo},\ and\
  \citenamefont {Novoselov}}]{gibertini2019magnetic}%
  \BibitemOpen
  \bibfield  {author} {\bibinfo {author} {\bibfnamefont {M.}~\bibnamefont
  {Gibertini}}, \bibinfo {author} {\bibfnamefont {M.}~\bibnamefont {Koperski}},
  \bibinfo {author} {\bibfnamefont {A.~F.}\ \bibnamefont {Morpurgo}}, \ and\
  \bibinfo {author} {\bibfnamefont {K.~S.}\ \bibnamefont {Novoselov}},\ }\href
  {\doibase https://doi.org/10.1038/s41565-019-0438-6} {\bibfield  {journal}
  {\bibinfo  {journal} {Nat. Nanotechnol.}\ }\textbf {\bibinfo {volume} {14}},\
  \bibinfo {pages} {408} (\bibinfo {year} {2019})}\BibitemShut {NoStop}%
\bibitem [{\citenamefont {Khan}\ \emph {et~al.}(2020)\citenamefont {Khan},
  \citenamefont {Tareen}, \citenamefont {Aslam}, \citenamefont {Wang},
  \citenamefont {Zhang}, \citenamefont {Mahmood}, \citenamefont {Ouyang},
  \citenamefont {Zhang},\ and\ \citenamefont {Guo}}]{khan2020recent}%
  \BibitemOpen
  \bibfield  {author} {\bibinfo {author} {\bibfnamefont {K.}~\bibnamefont
  {Khan}}, \bibinfo {author} {\bibfnamefont {A.~K.}\ \bibnamefont {Tareen}},
  \bibinfo {author} {\bibfnamefont {M.}~\bibnamefont {Aslam}}, \bibinfo
  {author} {\bibfnamefont {R.}~\bibnamefont {Wang}}, \bibinfo {author}
  {\bibfnamefont {Y.}~\bibnamefont {Zhang}}, \bibinfo {author} {\bibfnamefont
  {A.}~\bibnamefont {Mahmood}}, \bibinfo {author} {\bibfnamefont
  {Z.}~\bibnamefont {Ouyang}}, \bibinfo {author} {\bibfnamefont
  {H.}~\bibnamefont {Zhang}}, \ and\ \bibinfo {author} {\bibfnamefont
  {Z.}~\bibnamefont {Guo}},\ }\href {\doibase DOI: 10.1039/c9tc04187g}
  {\bibfield  {journal} {\bibinfo  {journal} {J. Mater. Chem. C}\ }\textbf
  {\bibinfo {volume} {8}},\ \bibinfo {pages} {387} (\bibinfo {year}
  {2020})}\BibitemShut {NoStop}%
\bibitem [{\citenamefont {Glavin}\ \emph {et~al.}(2020)\citenamefont {Glavin},
  \citenamefont {Rao}, \citenamefont {Varshney}, \citenamefont {Bianco},
  \citenamefont {Apte}, \citenamefont {Roy}, \citenamefont {Ringe},\ and\
  \citenamefont {Ajayan}}]{glavin2020emerging}%
  \BibitemOpen
  \bibfield  {author} {\bibinfo {author} {\bibfnamefont {N.~R.}\ \bibnamefont
  {Glavin}}, \bibinfo {author} {\bibfnamefont {R.}~\bibnamefont {Rao}},
  \bibinfo {author} {\bibfnamefont {V.}~\bibnamefont {Varshney}}, \bibinfo
  {author} {\bibfnamefont {E.}~\bibnamefont {Bianco}}, \bibinfo {author}
  {\bibfnamefont {A.}~\bibnamefont {Apte}}, \bibinfo {author} {\bibfnamefont
  {A.}~\bibnamefont {Roy}}, \bibinfo {author} {\bibfnamefont {E.}~\bibnamefont
  {Ringe}}, \ and\ \bibinfo {author} {\bibfnamefont {P.~M.}\ \bibnamefont
  {Ajayan}},\ }\href {\doibase https://doi.org/10.1002/adma.201904302}
  {\bibfield  {journal} {\bibinfo  {journal} {Adv. Mater.}\ }\textbf {\bibinfo
  {volume} {32}},\ \bibinfo {pages} {1904302} (\bibinfo {year}
  {2020})}\BibitemShut {NoStop}%
\bibitem [{\citenamefont {Kumbhakar}\ \emph {et~al.}(2023)\citenamefont
  {Kumbhakar}, \citenamefont {Jayan}, \citenamefont {Madhavikutty},
  \citenamefont {Sreeram}, \citenamefont {Appukuttan}, \citenamefont {Ito},\
  and\ \citenamefont {Tiwary}}]{kumbhakar2023prospective}%
  \BibitemOpen
  \bibfield  {author} {\bibinfo {author} {\bibfnamefont {P.}~\bibnamefont
  {Kumbhakar}}, \bibinfo {author} {\bibfnamefont {J.~S.}\ \bibnamefont
  {Jayan}}, \bibinfo {author} {\bibfnamefont {A.~S.}\ \bibnamefont
  {Madhavikutty}}, \bibinfo {author} {\bibfnamefont {P.}~\bibnamefont
  {Sreeram}}, \bibinfo {author} {\bibfnamefont {S.}~\bibnamefont {Appukuttan}},
  \bibinfo {author} {\bibfnamefont {T.}~\bibnamefont {Ito}}, \ and\ \bibinfo
  {author} {\bibfnamefont {C.~S.}\ \bibnamefont {Tiwary}},\ }\href
  {https://doi.org/10.1016/j.isci.2023.106671} {\bibfield  {journal} {\bibinfo
  {journal} {IScience}\ } (\bibinfo {year} {2023})}\BibitemShut {NoStop}%
\bibitem [{\citenamefont {Huang}\ \emph {et~al.}(2017)\citenamefont {Huang},
  \citenamefont {Clark}, \citenamefont {Navarro-Moratalla}, \citenamefont
  {Klein}, \citenamefont {Cheng}, \citenamefont {Seyler}, \citenamefont
  {Zhong}, \citenamefont {Schmidgall}, \citenamefont {McGuire}, \citenamefont
  {Cobden} \emph {et~al.}}]{huang2017layer}%
  \BibitemOpen
  \bibfield  {author} {\bibinfo {author} {\bibfnamefont {B.}~\bibnamefont
  {Huang}}, \bibinfo {author} {\bibfnamefont {G.}~\bibnamefont {Clark}},
  \bibinfo {author} {\bibfnamefont {E.}~\bibnamefont {Navarro-Moratalla}},
  \bibinfo {author} {\bibfnamefont {D.~R.}\ \bibnamefont {Klein}}, \bibinfo
  {author} {\bibfnamefont {R.}~\bibnamefont {Cheng}}, \bibinfo {author}
  {\bibfnamefont {K.~L.}\ \bibnamefont {Seyler}}, \bibinfo {author}
  {\bibfnamefont {D.}~\bibnamefont {Zhong}}, \bibinfo {author} {\bibfnamefont
  {E.}~\bibnamefont {Schmidgall}}, \bibinfo {author} {\bibfnamefont {M.~A.}\
  \bibnamefont {McGuire}}, \bibinfo {author} {\bibfnamefont {D.~H.}\
  \bibnamefont {Cobden}},  \emph {et~al.},\ }\href {\doibase
  https://doi.org/10.1038/nature22391} {\bibfield  {journal} {\bibinfo
  {journal} {Nature}\ }\textbf {\bibinfo {volume} {546}},\ \bibinfo {pages}
  {270} (\bibinfo {year} {2017})}\BibitemShut {NoStop}%
\bibitem [{\citenamefont {Gong}\ \emph {et~al.}(2017)\citenamefont {Gong},
  \citenamefont {Li}, \citenamefont {Li}, \citenamefont {Ji}, \citenamefont
  {Stern}, \citenamefont {Xia}, \citenamefont {Cao}, \citenamefont {Bao},
  \citenamefont {Wang}, \citenamefont {Wang} \emph
  {et~al.}}]{gong2017discovery}%
  \BibitemOpen
  \bibfield  {author} {\bibinfo {author} {\bibfnamefont {C.}~\bibnamefont
  {Gong}}, \bibinfo {author} {\bibfnamefont {L.}~\bibnamefont {Li}}, \bibinfo
  {author} {\bibfnamefont {Z.}~\bibnamefont {Li}}, \bibinfo {author}
  {\bibfnamefont {H.}~\bibnamefont {Ji}}, \bibinfo {author} {\bibfnamefont
  {A.}~\bibnamefont {Stern}}, \bibinfo {author} {\bibfnamefont
  {Y.}~\bibnamefont {Xia}}, \bibinfo {author} {\bibfnamefont {T.}~\bibnamefont
  {Cao}}, \bibinfo {author} {\bibfnamefont {W.}~\bibnamefont {Bao}}, \bibinfo
  {author} {\bibfnamefont {C.}~\bibnamefont {Wang}}, \bibinfo {author}
  {\bibfnamefont {Y.}~\bibnamefont {Wang}},  \emph {et~al.},\ }\href {\doibase
  https://doi.org/10.1038/nature22060} {\bibfield  {journal} {\bibinfo
  {journal} {Nature}\ }\textbf {\bibinfo {volume} {546}},\ \bibinfo {pages}
  {265} (\bibinfo {year} {2017})}\BibitemShut {NoStop}%
\bibitem [{\citenamefont {Li}\ and\ \citenamefont {Wu}(2017)}]{li2017binary}%
  \BibitemOpen
  \bibfield  {author} {\bibinfo {author} {\bibfnamefont {L.}~\bibnamefont
  {Li}}\ and\ \bibinfo {author} {\bibfnamefont {M.}~\bibnamefont {Wu}},\ }\href
  {https://pubs.acs.org/doi/abs/10.1021/acsnano.7b02756} {\bibfield  {journal}
  {\bibinfo  {journal} {ACS Nano}\ }\textbf {\bibinfo {volume} {11}},\ \bibinfo
  {pages} {6382} (\bibinfo {year} {2017})}\BibitemShut {NoStop}%
\bibitem [{\citenamefont {Stern}\ \emph {et~al.}(2021)\citenamefont {Stern},
  \citenamefont {Waschitz}, \citenamefont {Cao}, \citenamefont {Nevo},
  \citenamefont {Watanabe}, \citenamefont {Taniguchi}, \citenamefont {Sela},
  \citenamefont {Urbakh}, \citenamefont {Hod},\ and\ \citenamefont
  {Shalom}}]{stern2020interfacial}%
  \BibitemOpen
  \bibfield  {author} {\bibinfo {author} {\bibfnamefont {M.~V.}\ \bibnamefont
  {Stern}}, \bibinfo {author} {\bibfnamefont {Y.}~\bibnamefont {Waschitz}},
  \bibinfo {author} {\bibfnamefont {W.}~\bibnamefont {Cao}}, \bibinfo {author}
  {\bibfnamefont {I.}~\bibnamefont {Nevo}}, \bibinfo {author} {\bibfnamefont
  {K.}~\bibnamefont {Watanabe}}, \bibinfo {author} {\bibfnamefont
  {T.}~\bibnamefont {Taniguchi}}, \bibinfo {author} {\bibfnamefont
  {E.}~\bibnamefont {Sela}}, \bibinfo {author} {\bibfnamefont {M.}~\bibnamefont
  {Urbakh}}, \bibinfo {author} {\bibfnamefont {O.}~\bibnamefont {Hod}}, \ and\
  \bibinfo {author} {\bibfnamefont {M.~B.}\ \bibnamefont {Shalom}},\ }\href
  {\doibase 10.1126/science.abe8177} {\bibfield  {journal} {\bibinfo  {journal}
  {Science}\ }\textbf {\bibinfo {volume} {372}},\ \bibinfo {pages} {1462}
  (\bibinfo {year} {2021})}\BibitemShut {NoStop}%
\bibitem [{\citenamefont {Yasuda}\ \emph {et~al.}(2021)\citenamefont {Yasuda},
  \citenamefont {Wang}, \citenamefont {Watanabe}, \citenamefont {Taniguchi},\
  and\ \citenamefont {Jarillo-Herrero}}]{yasuda2021stacking}%
  \BibitemOpen
  \bibfield  {author} {\bibinfo {author} {\bibfnamefont {K.}~\bibnamefont
  {Yasuda}}, \bibinfo {author} {\bibfnamefont {X.}~\bibnamefont {Wang}},
  \bibinfo {author} {\bibfnamefont {K.}~\bibnamefont {Watanabe}}, \bibinfo
  {author} {\bibfnamefont {T.}~\bibnamefont {Taniguchi}}, \ and\ \bibinfo
  {author} {\bibfnamefont {P.}~\bibnamefont {Jarillo-Herrero}},\ }\href
  {https://doi.org/10.1126/science.abd3230} {\bibfield  {journal} {\bibinfo
  {journal} {Science}\ }\textbf {\bibinfo {volume} {372}},\ \bibinfo {pages}
  {1458} (\bibinfo {year} {2021})}\BibitemShut {NoStop}%
\bibitem [{\citenamefont {Mermin}\ and\ \citenamefont
  {Wagner}(1966)}]{mermin1966absence}%
  \BibitemOpen
  \bibfield  {author} {\bibinfo {author} {\bibfnamefont {N.~D.}\ \bibnamefont
  {Mermin}}\ and\ \bibinfo {author} {\bibfnamefont {H.}~\bibnamefont
  {Wagner}},\ }\href {\doibase 10.1103/PhysRevLett.17.1133} {\bibfield
  {journal} {\bibinfo  {journal} {Phys. Rev. Lett.}\ }\textbf {\bibinfo
  {volume} {17}},\ \bibinfo {pages} {1133} (\bibinfo {year}
  {1966})}\BibitemShut {NoStop}%
\bibitem [{\citenamefont {{{\v{S}}i{\v{s}}kins, Makars and Kurdi, Samer and
  Lee, Martin and Slotboom, Benjamin JM and Xing, Wenyu and Ma{\~n}as-Valero,
  Samuel and Coronado, Eugenio and Jia, Shuang and Han, Wei and van der Sar,
  Toeno and others}}(2022)}]{vsivskins2022nanomechanical}%
  \BibitemOpen
  \bibfield  {author} {\bibinfo {author} {\bibnamefont {{{\v{S}}i{\v{s}}kins,
  Makars and Kurdi, Samer and Lee, Martin and Slotboom, Benjamin JM and Xing,
  Wenyu and Ma{\~n}as-Valero, Samuel and Coronado, Eugenio and Jia, Shuang and
  Han, Wei and van der Sar, Toeno and others}}},\ }\href {\doibase
  https://doi.org/10.1038/s41699-022-00315-7} {\bibfield  {journal} {\bibinfo
  {journal} {npj 2D Mater. Appl.}\ }\textbf {\bibinfo {volume} {6}},\ \bibinfo
  {pages} {41} (\bibinfo {year} {2022})}\BibitemShut {NoStop}%
\bibitem [{\citenamefont {Lv}\ \emph {et~al.}(2019)\citenamefont {Lv},
  \citenamefont {Lu}, \citenamefont {Luo}, \citenamefont {Zhu},\ and\
  \citenamefont {Sun}}]{lv2019strain}%
  \BibitemOpen
  \bibfield  {author} {\bibinfo {author} {\bibfnamefont {H.~Y.}\ \bibnamefont
  {Lv}}, \bibinfo {author} {\bibfnamefont {W.~J.}\ \bibnamefont {Lu}}, \bibinfo
  {author} {\bibfnamefont {X.}~\bibnamefont {Luo}}, \bibinfo {author}
  {\bibfnamefont {X.~B.}\ \bibnamefont {Zhu}}, \ and\ \bibinfo {author}
  {\bibfnamefont {Y.~P.}\ \bibnamefont {Sun}},\ }\href {\doibase
  10.1103/PhysRevB.99.134416} {\bibfield  {journal} {\bibinfo  {journal} {Phys.
  Rev. B}\ }\textbf {\bibinfo {volume} {99}},\ \bibinfo {pages} {134416}
  (\bibinfo {year} {2019})}\BibitemShut {NoStop}%
\bibitem [{\citenamefont {Mart\'{\i}nez-Carracedo}\ \emph
  {et~al.}(2023{\natexlab{a}})\citenamefont {Mart\'{\i}nez-Carracedo},
  \citenamefont {Oroszl\'any}, \citenamefont {Garc\'{\i}a-Fuente},
  \citenamefont {Szunyogh},\ and\ \citenamefont
  {Ferrer}}]{martinez2023electrically}%
  \BibitemOpen
  \bibfield  {author} {\bibinfo {author} {\bibfnamefont {G.}~\bibnamefont
  {Mart\'{\i}nez-Carracedo}}, \bibinfo {author} {\bibfnamefont
  {L.}~\bibnamefont {Oroszl\'any}}, \bibinfo {author} {\bibfnamefont
  {A.}~\bibnamefont {Garc\'{\i}a-Fuente}}, \bibinfo {author} {\bibfnamefont
  {L.}~\bibnamefont {Szunyogh}}, \ and\ \bibinfo {author} {\bibfnamefont
  {J.}~\bibnamefont {Ferrer}},\ }\href {\doibase 10.1103/PhysRevB.107.035432}
  {\bibfield  {journal} {\bibinfo  {journal} {Phys. Rev. B}\ }\textbf {\bibinfo
  {volume} {107}},\ \bibinfo {pages} {035432} (\bibinfo {year}
  {2023}{\natexlab{a}})}\BibitemShut {NoStop}%
\bibitem [{\citenamefont {Wang}\ \emph {et~al.}(2022)\citenamefont {Wang},
  \citenamefont {Yasuda}, \citenamefont {Zhang}, \citenamefont {Liu},
  \citenamefont {Watanabe}, \citenamefont {Taniguchi}, \citenamefont {Hone},
  \citenamefont {Fu},\ and\ \citenamefont
  {Jarillo-Herrero}}]{wang2022interfacial}%
  \BibitemOpen
  \bibfield  {author} {\bibinfo {author} {\bibfnamefont {X.}~\bibnamefont
  {Wang}}, \bibinfo {author} {\bibfnamefont {K.}~\bibnamefont {Yasuda}},
  \bibinfo {author} {\bibfnamefont {Y.}~\bibnamefont {Zhang}}, \bibinfo
  {author} {\bibfnamefont {S.}~\bibnamefont {Liu}}, \bibinfo {author}
  {\bibfnamefont {K.}~\bibnamefont {Watanabe}}, \bibinfo {author}
  {\bibfnamefont {T.}~\bibnamefont {Taniguchi}}, \bibinfo {author}
  {\bibfnamefont {J.}~\bibnamefont {Hone}}, \bibinfo {author} {\bibfnamefont
  {L.}~\bibnamefont {Fu}}, \ and\ \bibinfo {author} {\bibfnamefont
  {P.}~\bibnamefont {Jarillo-Herrero}},\ }\href {\doibase
  https://doi.org/10.1038/s41565-021-01059-z} {\bibfield  {journal} {\bibinfo
  {journal} {Nat. Nanotechnol.}\ }\textbf {\bibinfo {volume} {17}},\ \bibinfo
  {pages} {367} (\bibinfo {year} {2022})}\BibitemShut {NoStop}%
\bibitem [{\citenamefont {Weston}\ \emph {et~al.}(2022)\citenamefont {Weston},
  \citenamefont {Castanon}, \citenamefont {Enaldiev}, \citenamefont {Ferreira},
  \citenamefont {Bhattacharjee}, \citenamefont {Xu}, \citenamefont
  {Corte-Le{\'o}n}, \citenamefont {Wu}, \citenamefont {Clark}, \citenamefont
  {Summerfield} \emph {et~al.}}]{weston2022interfacial}%
  \BibitemOpen
  \bibfield  {author} {\bibinfo {author} {\bibfnamefont {A.}~\bibnamefont
  {Weston}}, \bibinfo {author} {\bibfnamefont {E.~G.}\ \bibnamefont
  {Castanon}}, \bibinfo {author} {\bibfnamefont {V.}~\bibnamefont {Enaldiev}},
  \bibinfo {author} {\bibfnamefont {F.}~\bibnamefont {Ferreira}}, \bibinfo
  {author} {\bibfnamefont {S.}~\bibnamefont {Bhattacharjee}}, \bibinfo {author}
  {\bibfnamefont {S.}~\bibnamefont {Xu}}, \bibinfo {author} {\bibfnamefont
  {H.}~\bibnamefont {Corte-Le{\'o}n}}, \bibinfo {author} {\bibfnamefont
  {Z.}~\bibnamefont {Wu}}, \bibinfo {author} {\bibfnamefont {N.}~\bibnamefont
  {Clark}}, \bibinfo {author} {\bibfnamefont {A.}~\bibnamefont {Summerfield}},
  \emph {et~al.},\ }\href {\doibase https://doi.org/10.1038/s41565-022-01072-w}
  {\bibfield  {journal} {\bibinfo  {journal} {Nat. Nanotechnol.}\ }\textbf
  {\bibinfo {volume} {17}},\ \bibinfo {pages} {390} (\bibinfo {year}
  {2022})}\BibitemShut {NoStop}%
\bibitem [{\citenamefont {Ko}\ \emph {et~al.}(2023)\citenamefont {Ko},
  \citenamefont {Yuk}, \citenamefont {Engelke}, \citenamefont {Carr},
  \citenamefont {Kim}, \citenamefont {Park}, \citenamefont {Heo}, \citenamefont
  {Kim}, \citenamefont {Kim}, \citenamefont {Kim} \emph
  {et~al.}}]{ko2023operando}%
  \BibitemOpen
  \bibfield  {author} {\bibinfo {author} {\bibfnamefont {K.}~\bibnamefont
  {Ko}}, \bibinfo {author} {\bibfnamefont {A.}~\bibnamefont {Yuk}}, \bibinfo
  {author} {\bibfnamefont {R.}~\bibnamefont {Engelke}}, \bibinfo {author}
  {\bibfnamefont {S.}~\bibnamefont {Carr}}, \bibinfo {author} {\bibfnamefont
  {J.}~\bibnamefont {Kim}}, \bibinfo {author} {\bibfnamefont {D.}~\bibnamefont
  {Park}}, \bibinfo {author} {\bibfnamefont {H.}~\bibnamefont {Heo}}, \bibinfo
  {author} {\bibfnamefont {H.-M.}\ \bibnamefont {Kim}}, \bibinfo {author}
  {\bibfnamefont {S.-G.}\ \bibnamefont {Kim}}, \bibinfo {author} {\bibfnamefont
  {H.}~\bibnamefont {Kim}},  \emph {et~al.},\ }\href {\doibase
  https://doi.org/10.1038/s41563-023-01595-0} {\bibfield  {journal} {\bibinfo
  {journal} {Nat. Mater.}\ ,\ \bibinfo {pages} {1}} (\bibinfo {year}
  {2023})}\BibitemShut {NoStop}%
\bibitem [{\citenamefont {Molino}\ \emph {et~al.}(2023)\citenamefont {Molino},
  \citenamefont {Aggarwal}, \citenamefont {Enaldiev}, \citenamefont
  {Plumadore}, \citenamefont {I.~Fal{\'{}}~ko},\ and\ \citenamefont
  {Luican-Mayer}}]{molino2023ferroelectric}%
  \BibitemOpen
  \bibfield  {author} {\bibinfo {author} {\bibfnamefont {L.}~\bibnamefont
  {Molino}}, \bibinfo {author} {\bibfnamefont {L.}~\bibnamefont {Aggarwal}},
  \bibinfo {author} {\bibfnamefont {V.}~\bibnamefont {Enaldiev}}, \bibinfo
  {author} {\bibfnamefont {R.}~\bibnamefont {Plumadore}}, \bibinfo {author}
  {\bibfnamefont {V.}~\bibnamefont {I.~Fal{\'{}}~ko}}, \ and\ \bibinfo {author}
  {\bibfnamefont {A.}~\bibnamefont {Luican-Mayer}},\ }\href {\doibase
  https://doi.org/10.1002/adma.202207816} {\bibfield  {journal} {\bibinfo
  {journal} {Adv. Mater.}\ }\textbf {\bibinfo {volume} {35}},\ \bibinfo {pages}
  {2207816} (\bibinfo {year} {2023})}\BibitemShut {NoStop}%
\bibitem [{\citenamefont {Van~Winkle}\ \emph {et~al.}(2024)\citenamefont
  {Van~Winkle}, \citenamefont {Dowlatshahi}, \citenamefont {Khaloo},
  \citenamefont {Iyer}, \citenamefont {Craig}, \citenamefont {Dhall},
  \citenamefont {Taniguchi}, \citenamefont {Watanabe},\ and\ \citenamefont
  {Bediako}}]{van2024engineering}%
  \BibitemOpen
  \bibfield  {author} {\bibinfo {author} {\bibfnamefont {M.}~\bibnamefont
  {Van~Winkle}}, \bibinfo {author} {\bibfnamefont {N.}~\bibnamefont
  {Dowlatshahi}}, \bibinfo {author} {\bibfnamefont {N.}~\bibnamefont {Khaloo}},
  \bibinfo {author} {\bibfnamefont {M.}~\bibnamefont {Iyer}}, \bibinfo {author}
  {\bibfnamefont {I.~M.}\ \bibnamefont {Craig}}, \bibinfo {author}
  {\bibfnamefont {R.}~\bibnamefont {Dhall}}, \bibinfo {author} {\bibfnamefont
  {T.}~\bibnamefont {Taniguchi}}, \bibinfo {author} {\bibfnamefont
  {K.}~\bibnamefont {Watanabe}}, \ and\ \bibinfo {author} {\bibfnamefont
  {D.~K.}\ \bibnamefont {Bediako}},\ }\href {\doibase
  https://doi.org/10.1038/s41565-024-01642-0} {\bibfield  {journal} {\bibinfo
  {journal} {Nat. Nanotechnol.}\ ,\ \bibinfo {pages} {1}} (\bibinfo {year}
  {2024})}\BibitemShut {NoStop}%
\bibitem [{\citenamefont {Yasuda}\ \emph {et~al.}(2024)\citenamefont {Yasuda},
  \citenamefont {Zalys-Geller}, \citenamefont {Wang}, \citenamefont {Bennett},
  \citenamefont {Cheema}, \citenamefont {Watanabe}, \citenamefont {Taniguchi},
  \citenamefont {Kaxiras}, \citenamefont {Jarillo-Herrero},\ and\ \citenamefont
  {Ashoori}}]{yasuda2024ultrafast}%
  \BibitemOpen
  \bibfield  {author} {\bibinfo {author} {\bibfnamefont {K.}~\bibnamefont
  {Yasuda}}, \bibinfo {author} {\bibfnamefont {E.}~\bibnamefont
  {Zalys-Geller}}, \bibinfo {author} {\bibfnamefont {X.}~\bibnamefont {Wang}},
  \bibinfo {author} {\bibfnamefont {D.}~\bibnamefont {Bennett}}, \bibinfo
  {author} {\bibfnamefont {S.~S.}\ \bibnamefont {Cheema}}, \bibinfo {author}
  {\bibfnamefont {K.}~\bibnamefont {Watanabe}}, \bibinfo {author}
  {\bibfnamefont {T.}~\bibnamefont {Taniguchi}}, \bibinfo {author}
  {\bibfnamefont {E.}~\bibnamefont {Kaxiras}}, \bibinfo {author} {\bibfnamefont
  {P.}~\bibnamefont {Jarillo-Herrero}}, \ and\ \bibinfo {author} {\bibfnamefont
  {R.}~\bibnamefont {Ashoori}},\ }\href {\doibase
  https://doi.org/10.1126/science.adp3575} {\bibfield  {journal} {\bibinfo
  {journal} {Science}\ ,\ \bibinfo {pages} {eadp3575}} (\bibinfo {year}
  {2024})}\BibitemShut {NoStop}%
\bibitem [{\citenamefont {Bian}\ \emph {et~al.}(2024)\citenamefont {Bian},
  \citenamefont {He}, \citenamefont {Pan}, \citenamefont {Li}, \citenamefont
  {Cao}, \citenamefont {Meng}, \citenamefont {Chen}, \citenamefont {Liu},
  \citenamefont {Zhong}, \citenamefont {Li} \emph
  {et~al.}}]{bian2024developing}%
  \BibitemOpen
  \bibfield  {author} {\bibinfo {author} {\bibfnamefont {R.}~\bibnamefont
  {Bian}}, \bibinfo {author} {\bibfnamefont {R.}~\bibnamefont {He}}, \bibinfo
  {author} {\bibfnamefont {E.}~\bibnamefont {Pan}}, \bibinfo {author}
  {\bibfnamefont {Z.}~\bibnamefont {Li}}, \bibinfo {author} {\bibfnamefont
  {G.}~\bibnamefont {Cao}}, \bibinfo {author} {\bibfnamefont {P.}~\bibnamefont
  {Meng}}, \bibinfo {author} {\bibfnamefont {J.}~\bibnamefont {Chen}}, \bibinfo
  {author} {\bibfnamefont {Q.}~\bibnamefont {Liu}}, \bibinfo {author}
  {\bibfnamefont {Z.}~\bibnamefont {Zhong}}, \bibinfo {author} {\bibfnamefont
  {W.}~\bibnamefont {Li}},  \emph {et~al.},\ }\href {\doibase
  https://doi.org/10.1126/science.ado1744} {\bibfield  {journal} {\bibinfo
  {journal} {Science}\ ,\ \bibinfo {pages} {eado1744}} (\bibinfo {year}
  {2024})}\BibitemShut {NoStop}%
\bibitem [{\citenamefont {Bistritzer}\ and\ \citenamefont
  {MacDonald}(2011)}]{Bistritzer2011}%
  \BibitemOpen
  \bibfield  {author} {\bibinfo {author} {\bibfnamefont {R.}~\bibnamefont
  {Bistritzer}}\ and\ \bibinfo {author} {\bibfnamefont {A.~H.}\ \bibnamefont
  {MacDonald}},\ }\href {\doibase 10.1073/pnas.1108174108} {\bibfield
  {journal} {\bibinfo  {journal} {PNAS}\ }\textbf {\bibinfo {volume} {108}},\
  \bibinfo {pages} {12233} (\bibinfo {year} {2011})}\BibitemShut {NoStop}%
\bibitem [{\citenamefont {Cao}\ \emph {et~al.}(2018{\natexlab{a}})\citenamefont
  {Cao}, \citenamefont {Fatemi}, \citenamefont {Fang}, \citenamefont
  {Watanabe}, \citenamefont {Taniguchi}, \citenamefont {Kaxiras},\ and\
  \citenamefont {Jarillo-Herrero}}]{cao2018unconventional}%
  \BibitemOpen
  \bibfield  {author} {\bibinfo {author} {\bibfnamefont {Y.}~\bibnamefont
  {Cao}}, \bibinfo {author} {\bibfnamefont {V.}~\bibnamefont {Fatemi}},
  \bibinfo {author} {\bibfnamefont {S.}~\bibnamefont {Fang}}, \bibinfo {author}
  {\bibfnamefont {K.}~\bibnamefont {Watanabe}}, \bibinfo {author}
  {\bibfnamefont {T.}~\bibnamefont {Taniguchi}}, \bibinfo {author}
  {\bibfnamefont {E.}~\bibnamefont {Kaxiras}}, \ and\ \bibinfo {author}
  {\bibfnamefont {P.}~\bibnamefont {Jarillo-Herrero}},\ }\href {\doibase
  https://doi.org/10.1038/nature26160} {\bibfield  {journal} {\bibinfo
  {journal} {Nature}\ }\textbf {\bibinfo {volume} {556}},\ \bibinfo {pages}
  {43} (\bibinfo {year} {2018}{\natexlab{a}})}\BibitemShut {NoStop}%
\bibitem [{\citenamefont {Cao}\ \emph {et~al.}(2018{\natexlab{b}})\citenamefont
  {Cao}, \citenamefont {Fatemi}, \citenamefont {Demir}, \citenamefont {Fang},
  \citenamefont {Tomarken}, \citenamefont {Luo}, \citenamefont
  {Sanchez-Yamagishi}, \citenamefont {Watanabe}, \citenamefont {Taniguchi},
  \citenamefont {Kaxiras} \emph {et~al.}}]{cao2018correlated}%
  \BibitemOpen
  \bibfield  {author} {\bibinfo {author} {\bibfnamefont {Y.}~\bibnamefont
  {Cao}}, \bibinfo {author} {\bibfnamefont {V.}~\bibnamefont {Fatemi}},
  \bibinfo {author} {\bibfnamefont {A.}~\bibnamefont {Demir}}, \bibinfo
  {author} {\bibfnamefont {S.}~\bibnamefont {Fang}}, \bibinfo {author}
  {\bibfnamefont {S.~L.}\ \bibnamefont {Tomarken}}, \bibinfo {author}
  {\bibfnamefont {J.~Y.}\ \bibnamefont {Luo}}, \bibinfo {author} {\bibfnamefont
  {J.~D.}\ \bibnamefont {Sanchez-Yamagishi}}, \bibinfo {author} {\bibfnamefont
  {K.}~\bibnamefont {Watanabe}}, \bibinfo {author} {\bibfnamefont
  {T.}~\bibnamefont {Taniguchi}}, \bibinfo {author} {\bibfnamefont
  {E.}~\bibnamefont {Kaxiras}},  \emph {et~al.},\ }\href {\doibase
  https://doi.org/10.1038/nature26154} {\bibfield  {journal} {\bibinfo
  {journal} {Nature}\ }\textbf {\bibinfo {volume} {556}},\ \bibinfo {pages}
  {80} (\bibinfo {year} {2018}{\natexlab{b}})}\BibitemShut {NoStop}%
\bibitem [{\citenamefont {Bennett}\ \emph {et~al.}(2024)\citenamefont
  {Bennett}, \citenamefont {Larson}, \citenamefont {Sharma}, \citenamefont
  {Carr},\ and\ \citenamefont {Kaxiras}}]{bennett2024twisted}%
  \BibitemOpen
  \bibfield  {author} {\bibinfo {author} {\bibfnamefont {D.}~\bibnamefont
  {Bennett}}, \bibinfo {author} {\bibfnamefont {D.~T.}\ \bibnamefont {Larson}},
  \bibinfo {author} {\bibfnamefont {L.}~\bibnamefont {Sharma}}, \bibinfo
  {author} {\bibfnamefont {S.}~\bibnamefont {Carr}}, \ and\ \bibinfo {author}
  {\bibfnamefont {E.}~\bibnamefont {Kaxiras}},\ }\href {\doibase
  10.1103/PhysRevB.109.155422} {\bibfield  {journal} {\bibinfo  {journal}
  {Phys. Rev. B}\ }\textbf {\bibinfo {volume} {109}},\ \bibinfo {pages}
  {155422} (\bibinfo {year} {2024})}\BibitemShut {NoStop}%
\bibitem [{\citenamefont {Bennett}\ and\ \citenamefont
  {Remez}(2022)}]{bennett2022electrically}%
  \BibitemOpen
  \bibfield  {author} {\bibinfo {author} {\bibfnamefont {D.}~\bibnamefont
  {Bennett}}\ and\ \bibinfo {author} {\bibfnamefont {B.}~\bibnamefont
  {Remez}},\ }\href {\doibase https://doi.org/10.1038/s41699-021-00281-6}
  {\bibfield  {journal} {\bibinfo  {journal} {npj 2D Mater. Appl.}\ }\textbf
  {\bibinfo {volume} {6}},\ \bibinfo {pages} {1} (\bibinfo {year}
  {2022})}\BibitemShut {NoStop}%
\bibitem [{\citenamefont {Bennett}(2022)}]{bennett2022theory}%
  \BibitemOpen
  \bibfield  {author} {\bibinfo {author} {\bibfnamefont {D.}~\bibnamefont
  {Bennett}},\ }\href {\doibase https://doi.org/10.1103/PhysRevB.105.235445}
  {\bibfield  {journal} {\bibinfo  {journal} {Phys. Rev. B}\ }\textbf {\bibinfo
  {volume} {105}},\ \bibinfo {pages} {235445} (\bibinfo {year}
  {2022})}\BibitemShut {NoStop}%
\bibitem [{\citenamefont {Bennett}\ \emph
  {et~al.}(2023{\natexlab{a}})\citenamefont {Bennett}, \citenamefont
  {Chaudhary}, \citenamefont {Slager}, \citenamefont {Bousquet},\ and\
  \citenamefont {Ghosez}}]{bennett2023polar}%
  \BibitemOpen
  \bibfield  {author} {\bibinfo {author} {\bibfnamefont {D.}~\bibnamefont
  {Bennett}}, \bibinfo {author} {\bibfnamefont {G.}~\bibnamefont {Chaudhary}},
  \bibinfo {author} {\bibfnamefont {R.-J.}\ \bibnamefont {Slager}}, \bibinfo
  {author} {\bibfnamefont {E.}~\bibnamefont {Bousquet}}, \ and\ \bibinfo
  {author} {\bibfnamefont {P.}~\bibnamefont {Ghosez}},\ }\href {\doibase
  https://doi.org/10.1038/s41467-023-37337-8} {\bibfield  {journal} {\bibinfo
  {journal} {Nat. Commun.}\ }\textbf {\bibinfo {volume} {14}},\ \bibinfo
  {pages} {1629} (\bibinfo {year} {2023}{\natexlab{a}})}\BibitemShut {NoStop}%
\bibitem [{\citenamefont {Bennett}\ \emph
  {et~al.}(2023{\natexlab{b}})\citenamefont {Bennett}, \citenamefont
  {Jankowski}, \citenamefont {Chaudhary}, \citenamefont {Kaxiras},\ and\
  \citenamefont {Slager}}]{bennett2023theory}%
  \BibitemOpen
  \bibfield  {author} {\bibinfo {author} {\bibfnamefont {D.}~\bibnamefont
  {Bennett}}, \bibinfo {author} {\bibfnamefont {W.~J.}\ \bibnamefont
  {Jankowski}}, \bibinfo {author} {\bibfnamefont {G.}~\bibnamefont
  {Chaudhary}}, \bibinfo {author} {\bibfnamefont {E.}~\bibnamefont {Kaxiras}},
  \ and\ \bibinfo {author} {\bibfnamefont {R.-J.}\ \bibnamefont {Slager}},\
  }\href {\doibase 10.1103/PhysRevResearch.5.033216} {\bibfield  {journal}
  {\bibinfo  {journal} {Phys. Rev. Res.}\ }\textbf {\bibinfo {volume} {5}},\
  \bibinfo {pages} {033216} (\bibinfo {year} {2023}{\natexlab{b}})}\BibitemShut
  {NoStop}%
\bibitem [{\citenamefont {Jankowski}\ \emph {et~al.}(2024)\citenamefont
  {Jankowski}, \citenamefont {Bennett}, \citenamefont {Agarwal}, \citenamefont
  {Chaudhary},\ and\ \citenamefont {Slager}}]{jankowski2024polarization}%
  \BibitemOpen
  \bibfield  {author} {\bibinfo {author} {\bibfnamefont {W.~J.}\ \bibnamefont
  {Jankowski}}, \bibinfo {author} {\bibfnamefont {D.}~\bibnamefont {Bennett}},
  \bibinfo {author} {\bibfnamefont {A.}~\bibnamefont {Agarwal}}, \bibinfo
  {author} {\bibfnamefont {G.}~\bibnamefont {Chaudhary}}, \ and\ \bibinfo
  {author} {\bibfnamefont {R.-J.}\ \bibnamefont {Slager}},\ }\href
  {https://arxiv.org/abs/2404.16919} {\bibfield  {journal} {\bibinfo  {journal}
  {arXiv:2404.16919}\ } (\bibinfo {year} {2024})}\BibitemShut {NoStop}%
\bibitem [{\citenamefont {Vu}\ \emph {et~al.}(2024)\citenamefont {Vu},
  \citenamefont {Bennett}, \citenamefont {Pallewella}, \citenamefont {Uddin},
  \citenamefont {Xing}, \citenamefont {Zhao}, \citenamefont {Lee},
  \citenamefont {Mao}, \citenamefont {Muir}, \citenamefont {Jia}, \citenamefont
  {Davis}, \citenamefont {Watanabe}, \citenamefont {Taniguchi}, \citenamefont
  {Adam}, \citenamefont {Sharma}, \citenamefont {Fuhrer},\ and\ \citenamefont
  {Edmonds}}]{vu2024imaging}%
  \BibitemOpen
  \bibfield  {author} {\bibinfo {author} {\bibfnamefont {T.-H.-Y.}\
  \bibnamefont {Vu}}, \bibinfo {author} {\bibfnamefont {D.}~\bibnamefont
  {Bennett}}, \bibinfo {author} {\bibfnamefont {G.~N.}\ \bibnamefont
  {Pallewella}}, \bibinfo {author} {\bibfnamefont {M.~H.}\ \bibnamefont
  {Uddin}}, \bibinfo {author} {\bibfnamefont {K.}~\bibnamefont {Xing}},
  \bibinfo {author} {\bibfnamefont {W.}~\bibnamefont {Zhao}}, \bibinfo {author}
  {\bibfnamefont {S.~H.}\ \bibnamefont {Lee}}, \bibinfo {author} {\bibfnamefont
  {Z.}~\bibnamefont {Mao}}, \bibinfo {author} {\bibfnamefont {J.~B.}\
  \bibnamefont {Muir}}, \bibinfo {author} {\bibfnamefont {L.}~\bibnamefont
  {Jia}}, \bibinfo {author} {\bibfnamefont {J.~A.}\ \bibnamefont {Davis}},
  \bibinfo {author} {\bibfnamefont {K.}~\bibnamefont {Watanabe}}, \bibinfo
  {author} {\bibfnamefont {T.}~\bibnamefont {Taniguchi}}, \bibinfo {author}
  {\bibfnamefont {S.}~\bibnamefont {Adam}}, \bibinfo {author} {\bibfnamefont
  {P.}~\bibnamefont {Sharma}}, \bibinfo {author} {\bibfnamefont {M.~S.}\
  \bibnamefont {Fuhrer}}, \ and\ \bibinfo {author} {\bibfnamefont {M.~T.}\
  \bibnamefont {Edmonds}},\ }\href {https://arxiv.org/abs/2405.15126}
  {\bibfield  {journal} {\bibinfo  {journal} {arxiv:2405.15126}\ } (\bibinfo
  {year} {2024})}\BibitemShut {NoStop}%
\bibitem [{\citenamefont {Das}\ \emph {et~al.}(2019)\citenamefont {Das},
  \citenamefont {Tang}, \citenamefont {Hong}, \citenamefont {Gon{\c{c}}alves},
  \citenamefont {McCarter}, \citenamefont {Klewe}, \citenamefont {Nguyen},
  \citenamefont {G{\'o}mez-Ortiz}, \citenamefont {Shafer}, \citenamefont
  {Arenholz} \emph {et~al.}}]{das2019observation}%
  \BibitemOpen
  \bibfield  {author} {\bibinfo {author} {\bibfnamefont {S.}~\bibnamefont
  {Das}}, \bibinfo {author} {\bibfnamefont {Y.}~\bibnamefont {Tang}}, \bibinfo
  {author} {\bibfnamefont {Z.}~\bibnamefont {Hong}}, \bibinfo {author}
  {\bibfnamefont {M.}~\bibnamefont {Gon{\c{c}}alves}}, \bibinfo {author}
  {\bibfnamefont {M.}~\bibnamefont {McCarter}}, \bibinfo {author}
  {\bibfnamefont {C.}~\bibnamefont {Klewe}}, \bibinfo {author} {\bibfnamefont
  {K.}~\bibnamefont {Nguyen}}, \bibinfo {author} {\bibfnamefont
  {F.}~\bibnamefont {G{\'o}mez-Ortiz}}, \bibinfo {author} {\bibfnamefont
  {P.}~\bibnamefont {Shafer}}, \bibinfo {author} {\bibfnamefont
  {E.}~\bibnamefont {Arenholz}},  \emph {et~al.},\ }\href {\doibase
  https://doi.org/10.1038/s41586-019-1092-8} {\bibfield  {journal} {\bibinfo
  {journal} {Nature}\ }\textbf {\bibinfo {volume} {568}},\ \bibinfo {pages}
  {368} (\bibinfo {year} {2019})}\BibitemShut {NoStop}%
\bibitem [{\citenamefont {Han}\ \emph {et~al.}(2022)\citenamefont {Han},
  \citenamefont {Addiego}, \citenamefont {Prokhorenko}, \citenamefont {Wang},
  \citenamefont {Fu}, \citenamefont {Nahas}, \citenamefont {Yan}, \citenamefont
  {Cai}, \citenamefont {Wei}, \citenamefont {Fang} \emph
  {et~al.}}]{han2022high}%
  \BibitemOpen
  \bibfield  {author} {\bibinfo {author} {\bibfnamefont {L.}~\bibnamefont
  {Han}}, \bibinfo {author} {\bibfnamefont {C.}~\bibnamefont {Addiego}},
  \bibinfo {author} {\bibfnamefont {S.}~\bibnamefont {Prokhorenko}}, \bibinfo
  {author} {\bibfnamefont {M.}~\bibnamefont {Wang}}, \bibinfo {author}
  {\bibfnamefont {H.}~\bibnamefont {Fu}}, \bibinfo {author} {\bibfnamefont
  {Y.}~\bibnamefont {Nahas}}, \bibinfo {author} {\bibfnamefont
  {X.}~\bibnamefont {Yan}}, \bibinfo {author} {\bibfnamefont {S.}~\bibnamefont
  {Cai}}, \bibinfo {author} {\bibfnamefont {T.}~\bibnamefont {Wei}}, \bibinfo
  {author} {\bibfnamefont {Y.}~\bibnamefont {Fang}},  \emph {et~al.},\ }\href
  {\doibase https://doi.org/10.1038/s41586-021-04338-w} {\bibfield  {journal}
  {\bibinfo  {journal} {Nature}\ }\textbf {\bibinfo {volume} {603}},\ \bibinfo
  {pages} {63} (\bibinfo {year} {2022})}\BibitemShut {NoStop}%
\bibitem [{\citenamefont {Junquera}\ \emph {et~al.}(2023)\citenamefont
  {Junquera}, \citenamefont {Nahas}, \citenamefont {Prokhorenko}, \citenamefont
  {Bellaiche}, \citenamefont {\'I\~niguez}, \citenamefont {Schlom},
  \citenamefont {Chen}, \citenamefont {Salahuddin}, \citenamefont {Muller},
  \citenamefont {Martin},\ and\ \citenamefont
  {Ramesh}}]{junquera2023topologicaly}%
  \BibitemOpen
  \bibfield  {author} {\bibinfo {author} {\bibfnamefont {J.}~\bibnamefont
  {Junquera}}, \bibinfo {author} {\bibfnamefont {Y.}~\bibnamefont {Nahas}},
  \bibinfo {author} {\bibfnamefont {S.}~\bibnamefont {Prokhorenko}}, \bibinfo
  {author} {\bibfnamefont {L.}~\bibnamefont {Bellaiche}}, \bibinfo {author}
  {\bibfnamefont {J.}~\bibnamefont {\'I\~niguez}}, \bibinfo {author}
  {\bibfnamefont {D.~G.}\ \bibnamefont {Schlom}}, \bibinfo {author}
  {\bibfnamefont {L.-Q.}\ \bibnamefont {Chen}}, \bibinfo {author}
  {\bibfnamefont {S.}~\bibnamefont {Salahuddin}}, \bibinfo {author}
  {\bibfnamefont {D.~A.}\ \bibnamefont {Muller}}, \bibinfo {author}
  {\bibfnamefont {L.~W.}\ \bibnamefont {Martin}}, \ and\ \bibinfo {author}
  {\bibfnamefont {R.}~\bibnamefont {Ramesh}},\ }\href {\doibase
  10.1103/RevModPhys.95.025001} {\bibfield  {journal} {\bibinfo  {journal}
  {Rev. Mod. Phys.}\ }\textbf {\bibinfo {volume} {95}},\ \bibinfo {pages}
  {025001} (\bibinfo {year} {2023})}\BibitemShut {NoStop}%
\bibitem [{\citenamefont {S{\'a}nchez-Santolino}\ \emph
  {et~al.}(2024)\citenamefont {S{\'a}nchez-Santolino}, \citenamefont {Rouco},
  \citenamefont {Puebla}, \citenamefont {Aramberri}, \citenamefont {Zamora},
  \citenamefont {Cabero}, \citenamefont {Cuellar}, \citenamefont {Munuera},
  \citenamefont {Mompean}, \citenamefont {Garcia-Hernandez} \emph
  {et~al.}}]{sanchez20242d}%
  \BibitemOpen
  \bibfield  {author} {\bibinfo {author} {\bibfnamefont {G.}~\bibnamefont
  {S{\'a}nchez-Santolino}}, \bibinfo {author} {\bibfnamefont {V.}~\bibnamefont
  {Rouco}}, \bibinfo {author} {\bibfnamefont {S.}~\bibnamefont {Puebla}},
  \bibinfo {author} {\bibfnamefont {H.}~\bibnamefont {Aramberri}}, \bibinfo
  {author} {\bibfnamefont {V.}~\bibnamefont {Zamora}}, \bibinfo {author}
  {\bibfnamefont {M.}~\bibnamefont {Cabero}}, \bibinfo {author} {\bibfnamefont
  {F.}~\bibnamefont {Cuellar}}, \bibinfo {author} {\bibfnamefont
  {C.}~\bibnamefont {Munuera}}, \bibinfo {author} {\bibfnamefont
  {F.}~\bibnamefont {Mompean}}, \bibinfo {author} {\bibfnamefont
  {M.}~\bibnamefont {Garcia-Hernandez}},  \emph {et~al.},\ }\href {\doibase
  https://doi.org/10.1038/s41586-023-06978-6} {\bibfield  {journal} {\bibinfo
  {journal} {Nature}\ }\textbf {\bibinfo {volume} {626}},\ \bibinfo {pages}
  {529} (\bibinfo {year} {2024})}\BibitemShut {NoStop}%
\bibitem [{\citenamefont {Song}\ \emph {et~al.}(2021)\citenamefont {Song},
  \citenamefont {Sun}, \citenamefont {Anderson}, \citenamefont {Wang},
  \citenamefont {Qian}, \citenamefont {Taniguchi}, \citenamefont {Watanabe},
  \citenamefont {McGuire}, \citenamefont {St{\"o}hr}, \citenamefont {Xiao}
  \emph {et~al.}}]{song2021direct}%
  \BibitemOpen
  \bibfield  {author} {\bibinfo {author} {\bibfnamefont {T.}~\bibnamefont
  {Song}}, \bibinfo {author} {\bibfnamefont {Q.-C.}\ \bibnamefont {Sun}},
  \bibinfo {author} {\bibfnamefont {E.}~\bibnamefont {Anderson}}, \bibinfo
  {author} {\bibfnamefont {C.}~\bibnamefont {Wang}}, \bibinfo {author}
  {\bibfnamefont {J.}~\bibnamefont {Qian}}, \bibinfo {author} {\bibfnamefont
  {T.}~\bibnamefont {Taniguchi}}, \bibinfo {author} {\bibfnamefont
  {K.}~\bibnamefont {Watanabe}}, \bibinfo {author} {\bibfnamefont {M.~A.}\
  \bibnamefont {McGuire}}, \bibinfo {author} {\bibfnamefont {R.}~\bibnamefont
  {St{\"o}hr}}, \bibinfo {author} {\bibfnamefont {D.}~\bibnamefont {Xiao}},
  \emph {et~al.},\ }\href {\doibase https://doi.org/10.1126/science.abj7478}
  {\bibfield  {journal} {\bibinfo  {journal} {Science}\ }\textbf {\bibinfo
  {volume} {374}},\ \bibinfo {pages} {1140} (\bibinfo {year}
  {2021})}\BibitemShut {NoStop}%
\bibitem [{\citenamefont {Fumega}\ and\ \citenamefont
  {Lado}(2023)}]{fumega2023moire}%
  \BibitemOpen
  \bibfield  {author} {\bibinfo {author} {\bibfnamefont {A.~O.}\ \bibnamefont
  {Fumega}}\ and\ \bibinfo {author} {\bibfnamefont {J.~L.}\ \bibnamefont
  {Lado}},\ }\href {\doibase https://doi.org/10.1088/2053-1583/acc671}
  {\bibfield  {journal} {\bibinfo  {journal} {2D Mater.}\ }\textbf {\bibinfo
  {volume} {10}},\ \bibinfo {pages} {025026} (\bibinfo {year}
  {2023})}\BibitemShut {NoStop}%
\bibitem [{\citenamefont {Spaldin}\ and\ \citenamefont
  {Fiebig}(2005)}]{spaldin2005renaissance}%
  \BibitemOpen
  \bibfield  {author} {\bibinfo {author} {\bibfnamefont {N.~A.}\ \bibnamefont
  {Spaldin}}\ and\ \bibinfo {author} {\bibfnamefont {M.}~\bibnamefont
  {Fiebig}},\ }\href {\doibase https://doi.org/10.1126/science.1113357}
  {\bibfield  {journal} {\bibinfo  {journal} {Science}\ }\textbf {\bibinfo
  {volume} {309}},\ \bibinfo {pages} {391} (\bibinfo {year}
  {2005})}\BibitemShut {NoStop}%
\bibitem [{\citenamefont {Spaldin}\ and\ \citenamefont
  {Ramesh}(2019)}]{spaldin2019advances}%
  \BibitemOpen
  \bibfield  {author} {\bibinfo {author} {\bibfnamefont {N.~A.}\ \bibnamefont
  {Spaldin}}\ and\ \bibinfo {author} {\bibfnamefont {R.}~\bibnamefont
  {Ramesh}},\ }\href {\doibase https://doi.org/10.1038/s41563-018-0275-2}
  {\bibfield  {journal} {\bibinfo  {journal} {Nat. Mater.}\ }\textbf {\bibinfo
  {volume} {18}},\ \bibinfo {pages} {203} (\bibinfo {year} {2019})}\BibitemShut
  {NoStop}%
\bibitem [{\citenamefont {Wang}\ \emph {et~al.}(2003)\citenamefont {Wang},
  \citenamefont {Neaton}, \citenamefont {Zheng}, \citenamefont {Nagarajan},
  \citenamefont {Ogale}, \citenamefont {Liu}, \citenamefont {Viehland},
  \citenamefont {Vaithyanathan}, \citenamefont {Schlom}, \citenamefont
  {Waghmare} \emph {et~al.}}]{wang2003epitaxial}%
  \BibitemOpen
  \bibfield  {author} {\bibinfo {author} {\bibfnamefont {J.}~\bibnamefont
  {Wang}}, \bibinfo {author} {\bibfnamefont {J.}~\bibnamefont {Neaton}},
  \bibinfo {author} {\bibfnamefont {H.}~\bibnamefont {Zheng}}, \bibinfo
  {author} {\bibfnamefont {V.}~\bibnamefont {Nagarajan}}, \bibinfo {author}
  {\bibfnamefont {S.}~\bibnamefont {Ogale}}, \bibinfo {author} {\bibfnamefont
  {B.}~\bibnamefont {Liu}}, \bibinfo {author} {\bibfnamefont {D.}~\bibnamefont
  {Viehland}}, \bibinfo {author} {\bibfnamefont {V.}~\bibnamefont
  {Vaithyanathan}}, \bibinfo {author} {\bibfnamefont {D.}~\bibnamefont
  {Schlom}}, \bibinfo {author} {\bibfnamefont {U.}~\bibnamefont {Waghmare}},
  \emph {et~al.},\ }\href {\doibase https://doi.org/10.1126/science.1080615}
  {\bibfield  {journal} {\bibinfo  {journal} {Science}\ }\textbf {\bibinfo
  {volume} {299}},\ \bibinfo {pages} {1719} (\bibinfo {year}
  {2003})}\BibitemShut {NoStop}%
\bibitem [{\citenamefont {Liu}\ \emph {et~al.}(2020)\citenamefont {Liu},
  \citenamefont {Wang}, \citenamefont {Wu}, \citenamefont {Chen}, \citenamefont
  {Wan}, \citenamefont {Wen}, \citenamefont {Yang}, \citenamefont {Liu},
  \citenamefont {Song},\ and\ \citenamefont {Xie}}]{liu2020vapor}%
  \BibitemOpen
  \bibfield  {author} {\bibinfo {author} {\bibfnamefont {H.}~\bibnamefont
  {Liu}}, \bibinfo {author} {\bibfnamefont {X.}~\bibnamefont {Wang}}, \bibinfo
  {author} {\bibfnamefont {J.}~\bibnamefont {Wu}}, \bibinfo {author}
  {\bibfnamefont {Y.}~\bibnamefont {Chen}}, \bibinfo {author} {\bibfnamefont
  {J.}~\bibnamefont {Wan}}, \bibinfo {author} {\bibfnamefont {R.}~\bibnamefont
  {Wen}}, \bibinfo {author} {\bibfnamefont {J.}~\bibnamefont {Yang}}, \bibinfo
  {author} {\bibfnamefont {Y.}~\bibnamefont {Liu}}, \bibinfo {author}
  {\bibfnamefont {Z.}~\bibnamefont {Song}}, \ and\ \bibinfo {author}
  {\bibfnamefont {L.}~\bibnamefont {Xie}},\ }\href {\doibase
  https://doi.org/10.1021/acsnano.0c04499} {\bibfield  {journal} {\bibinfo
  {journal} {ACS nano}\ }\textbf {\bibinfo {volume} {14}},\ \bibinfo {pages}
  {10544} (\bibinfo {year} {2020})}\BibitemShut {NoStop}%
\bibitem [{\citenamefont {Ju}\ \emph {et~al.}(2021)\citenamefont {Ju},
  \citenamefont {Lee}, \citenamefont {Kim}, \citenamefont {Choi}, \citenamefont
  {Roh}, \citenamefont {Son}, \citenamefont {Park}, \citenamefont {Kim},
  \citenamefont {Jung}, \citenamefont {Kim} \emph {et~al.}}]{ju2021possible}%
  \BibitemOpen
  \bibfield  {author} {\bibinfo {author} {\bibfnamefont {H.}~\bibnamefont
  {Ju}}, \bibinfo {author} {\bibfnamefont {Y.}~\bibnamefont {Lee}}, \bibinfo
  {author} {\bibfnamefont {K.-T.}\ \bibnamefont {Kim}}, \bibinfo {author}
  {\bibfnamefont {I.~H.}\ \bibnamefont {Choi}}, \bibinfo {author}
  {\bibfnamefont {C.~J.}\ \bibnamefont {Roh}}, \bibinfo {author} {\bibfnamefont
  {S.}~\bibnamefont {Son}}, \bibinfo {author} {\bibfnamefont {P.}~\bibnamefont
  {Park}}, \bibinfo {author} {\bibfnamefont {J.~H.}\ \bibnamefont {Kim}},
  \bibinfo {author} {\bibfnamefont {T.~S.}\ \bibnamefont {Jung}}, \bibinfo
  {author} {\bibfnamefont {J.~H.}\ \bibnamefont {Kim}},  \emph {et~al.},\
  }\href {\doibase https://doi.org/10.1021/acs.nanolett.1c01095} {\bibfield
  {journal} {\bibinfo  {journal} {Nano Lett.}\ }\textbf {\bibinfo {volume}
  {21}},\ \bibinfo {pages} {5126} (\bibinfo {year} {2021})}\BibitemShut
  {NoStop}%
\bibitem [{\citenamefont {Jenkins}\ \emph {et~al.}(2022)\citenamefont
  {Jenkins}, \citenamefont {Rozsa}, \citenamefont {Atxitia}, \citenamefont
  {Evans}, \citenamefont {Novoselov},\ and\ \citenamefont
  {Santos}}]{Santos2022}%
  \BibitemOpen
  \bibfield  {author} {\bibinfo {author} {\bibfnamefont {S.}~\bibnamefont
  {Jenkins}}, \bibinfo {author} {\bibfnamefont {L.}~\bibnamefont {Rozsa}},
  \bibinfo {author} {\bibfnamefont {U.}~\bibnamefont {Atxitia}}, \bibinfo
  {author} {\bibfnamefont {R.~L.~F.}\ \bibnamefont {Evans}}, \bibinfo {author}
  {\bibfnamefont {K.~S.}\ \bibnamefont {Novoselov}}, \ and\ \bibinfo {author}
  {\bibfnamefont {E.~J.~G.}\ \bibnamefont {Santos}},\ }\href {\doibase
  doi={https://doi.org/10.1038/s41467-022-34389-0}} {\bibfield  {journal}
  {\bibinfo  {journal} {Nat. Commun.}\ }\textbf {\bibinfo {volume} {13}},\
  \bibinfo {pages} {6917} (\bibinfo {year} {2022})}\BibitemShut {NoStop}%
\bibitem [{\citenamefont {Song}\ \emph {et~al.}(2022)\citenamefont {Song},
  \citenamefont {Occhialini}, \citenamefont {Erge{\c{c}}en}, \citenamefont
  {Ilyas}, \citenamefont {Amoroso}, \citenamefont {Barone}, \citenamefont
  {Kapeghian}, \citenamefont {Watanabe}, \citenamefont {Taniguchi},
  \citenamefont {Botana} \emph {et~al.}}]{song2022evidence}%
  \BibitemOpen
  \bibfield  {author} {\bibinfo {author} {\bibfnamefont {Q.}~\bibnamefont
  {Song}}, \bibinfo {author} {\bibfnamefont {C.~A.}\ \bibnamefont
  {Occhialini}}, \bibinfo {author} {\bibfnamefont {E.}~\bibnamefont
  {Erge{\c{c}}en}}, \bibinfo {author} {\bibfnamefont {B.}~\bibnamefont
  {Ilyas}}, \bibinfo {author} {\bibfnamefont {D.}~\bibnamefont {Amoroso}},
  \bibinfo {author} {\bibfnamefont {P.}~\bibnamefont {Barone}}, \bibinfo
  {author} {\bibfnamefont {J.}~\bibnamefont {Kapeghian}}, \bibinfo {author}
  {\bibfnamefont {K.}~\bibnamefont {Watanabe}}, \bibinfo {author}
  {\bibfnamefont {T.}~\bibnamefont {Taniguchi}}, \bibinfo {author}
  {\bibfnamefont {A.~S.}\ \bibnamefont {Botana}},  \emph {et~al.},\ }\href
  {\doibase https://doi.org/10.1038/s41586-021-04337-x} {\bibfield  {journal}
  {\bibinfo  {journal} {Nature}\ }\textbf {\bibinfo {volume} {602}},\ \bibinfo
  {pages} {601} (\bibinfo {year} {2022})}\BibitemShut {NoStop}%
\bibitem [{\citenamefont {Mostovoy}(2006)}]{mostovoy2006ferroelectricity}%
  \BibitemOpen
  \bibfield  {author} {\bibinfo {author} {\bibfnamefont {M.}~\bibnamefont
  {Mostovoy}},\ }\href {\doibase 10.1103/PhysRevLett.96.067601} {\bibfield
  {journal} {\bibinfo  {journal} {Phys. Rev. Lett.}\ }\textbf {\bibinfo
  {volume} {96}},\ \bibinfo {pages} {067601} (\bibinfo {year}
  {2006})}\BibitemShut {NoStop}%
\bibitem [{\citenamefont {Xiang}\ \emph {et~al.}(2011)\citenamefont {Xiang},
  \citenamefont {Kan}, \citenamefont {Zhang}, \citenamefont {Whangbo},\ and\
  \citenamefont {Gong}}]{xiang2011general}%
  \BibitemOpen
  \bibfield  {author} {\bibinfo {author} {\bibfnamefont {H.~J.}\ \bibnamefont
  {Xiang}}, \bibinfo {author} {\bibfnamefont {E.~J.}\ \bibnamefont {Kan}},
  \bibinfo {author} {\bibfnamefont {Y.}~\bibnamefont {Zhang}}, \bibinfo
  {author} {\bibfnamefont {M.-H.}\ \bibnamefont {Whangbo}}, \ and\ \bibinfo
  {author} {\bibfnamefont {X.~G.}\ \bibnamefont {Gong}},\ }\href {\doibase
  10.1103/PhysRevLett.107.157202} {\bibfield  {journal} {\bibinfo  {journal}
  {Phys. Rev. Lett.}\ }\textbf {\bibinfo {volume} {107}},\ \bibinfo {pages}
  {157202} (\bibinfo {year} {2011})}\BibitemShut {NoStop}%
\bibitem [{\citenamefont {Hohenberg}\ and\ \citenamefont
  {Krekhov}(2015)}]{hohenberg2015introduction}%
  \BibitemOpen
  \bibfield  {author} {\bibinfo {author} {\bibfnamefont {P.~C.}\ \bibnamefont
  {Hohenberg}}\ and\ \bibinfo {author} {\bibfnamefont {A.~P.}\ \bibnamefont
  {Krekhov}},\ }\href {\doibase https://doi.org/10.1016/j.physrep.2015.01.001}
  {\bibfield  {journal} {\bibinfo  {journal} {Phys. Rep.}\ }\textbf {\bibinfo
  {volume} {572}},\ \bibinfo {pages} {1} (\bibinfo {year} {2015})}\BibitemShut
  {NoStop}%
\bibitem [{\citenamefont {Fumega}\ and\ \citenamefont
  {Lado}(2022)}]{fumega2022microscopic}%
  \BibitemOpen
  \bibfield  {author} {\bibinfo {author} {\bibfnamefont {A.~O.}\ \bibnamefont
  {Fumega}}\ and\ \bibinfo {author} {\bibfnamefont {J.}~\bibnamefont {Lado}},\
  }\href {\doibase 10.1088/2053-1583/ac4e9d} {\bibfield  {journal} {\bibinfo
  {journal} {2D Mater.}\ }\textbf {\bibinfo {volume} {9}},\ \bibinfo {pages}
  {025010} (\bibinfo {year} {2022})}\BibitemShut {NoStop}%
\bibitem [{\citenamefont {Sivadas}\ \emph {et~al.}(2018)\citenamefont
  {Sivadas}, \citenamefont {Okamoto}, \citenamefont {Xu}, \citenamefont
  {Fennie},\ and\ \citenamefont {Xiao}}]{sivadas2018stacking}%
  \BibitemOpen
  \bibfield  {author} {\bibinfo {author} {\bibfnamefont {N.}~\bibnamefont
  {Sivadas}}, \bibinfo {author} {\bibfnamefont {S.}~\bibnamefont {Okamoto}},
  \bibinfo {author} {\bibfnamefont {X.}~\bibnamefont {Xu}}, \bibinfo {author}
  {\bibfnamefont {C.~J.}\ \bibnamefont {Fennie}}, \ and\ \bibinfo {author}
  {\bibfnamefont {D.}~\bibnamefont {Xiao}},\ }\href {\doibase
  https://doi.org/10.1021/acs.nanolett.8b03321} {\bibfield  {journal} {\bibinfo
   {journal} {Nano Lett.}\ }\textbf {\bibinfo {volume} {18}},\ \bibinfo {pages}
  {7658} (\bibinfo {year} {2018})}\BibitemShut {NoStop}%
\bibitem [{\citenamefont {Ji}\ \emph {et~al.}(2023)\citenamefont {Ji},
  \citenamefont {Yu}, \citenamefont {Xu},\ and\ \citenamefont
  {Xiang}}]{ji2023general}%
  \BibitemOpen
  \bibfield  {author} {\bibinfo {author} {\bibfnamefont {J.}~\bibnamefont
  {Ji}}, \bibinfo {author} {\bibfnamefont {G.}~\bibnamefont {Yu}}, \bibinfo
  {author} {\bibfnamefont {C.}~\bibnamefont {Xu}}, \ and\ \bibinfo {author}
  {\bibfnamefont {H.~J.}\ \bibnamefont {Xiang}},\ }\href {\doibase
  10.1103/PhysRevLett.130.146801} {\bibfield  {journal} {\bibinfo  {journal}
  {Phys. Rev. Lett.}\ }\textbf {\bibinfo {volume} {130}},\ \bibinfo {pages}
  {146801} (\bibinfo {year} {2023})}\BibitemShut {NoStop}%
\bibitem [{\citenamefont {Xun}\ \emph {et~al.}(2024)\citenamefont {Xun},
  \citenamefont {Wu}, \citenamefont {Sun}, \citenamefont {Zhang}, \citenamefont
  {Wu},\ and\ \citenamefont {Li}}]{xun2024coexisting}%
  \BibitemOpen
  \bibfield  {author} {\bibinfo {author} {\bibfnamefont {W.}~\bibnamefont
  {Xun}}, \bibinfo {author} {\bibfnamefont {C.}~\bibnamefont {Wu}}, \bibinfo
  {author} {\bibfnamefont {H.}~\bibnamefont {Sun}}, \bibinfo {author}
  {\bibfnamefont {W.}~\bibnamefont {Zhang}}, \bibinfo {author} {\bibfnamefont
  {Y.-Z.}\ \bibnamefont {Wu}}, \ and\ \bibinfo {author} {\bibfnamefont
  {P.}~\bibnamefont {Li}},\ }\href {\doibase
  https://doi.org/10.1021/acs.nanolett.4c00597} {\bibfield  {journal} {\bibinfo
   {journal} {Nano Lett.}\ }\textbf {\bibinfo {volume} {24}},\ \bibinfo {pages}
  {3541} (\bibinfo {year} {2024})}\BibitemShut {NoStop}%
\bibitem [{SM()}]{SM}%
  \BibitemOpen
  \href@noop {} {}\bibinfo {note} {See Supplemental Material [url] for details
  of first-principles calculations and additional results, which includes
  Refs.~\cite{siesta,norm_conserving,psml,pseudodojo,papior2018simple,gonze2009abinit,mp,pbe,grimme2010consistent,dipole_correction_1,rutter2018c2x}.}\BibitemShut
  {Stop}%
\bibitem [{\citenamefont {Liechtenstein}\ \emph {et~al.}(1987)\citenamefont
  {Liechtenstein}, \citenamefont {Katsnelson}, \citenamefont {Antropov},\ and\
  \citenamefont {Gubanov}}]{liechtenstein1987local}%
  \BibitemOpen
  \bibfield  {author} {\bibinfo {author} {\bibfnamefont {A.~I.}\ \bibnamefont
  {Liechtenstein}}, \bibinfo {author} {\bibfnamefont {M.}~\bibnamefont
  {Katsnelson}}, \bibinfo {author} {\bibfnamefont {V.}~\bibnamefont
  {Antropov}}, \ and\ \bibinfo {author} {\bibfnamefont {V.}~\bibnamefont
  {Gubanov}},\ }\href {\doibase https://doi.org/10.1016/0304-8853(87)90721-9}
  {\bibfield  {journal} {\bibinfo  {journal} {J. Magn. Magn. Mater.}\ }\textbf
  {\bibinfo {volume} {67}},\ \bibinfo {pages} {65} (\bibinfo {year}
  {1987})}\BibitemShut {NoStop}%
\bibitem [{\citenamefont {Mart\'{\i}nez-Carracedo}\ \emph
  {et~al.}(2023{\natexlab{b}})\citenamefont {Mart\'{\i}nez-Carracedo},
  \citenamefont {Oroszl\'any}, \citenamefont {Garc\'{\i}a-Fuente},
  \citenamefont {Ny\'ari}, \citenamefont {Udvardi}, \citenamefont {Szunyogh},\
  and\ \citenamefont {Ferrer}}]{grogu}%
  \BibitemOpen
  \bibfield  {author} {\bibinfo {author} {\bibfnamefont {G.}~\bibnamefont
  {Mart\'{\i}nez-Carracedo}}, \bibinfo {author} {\bibfnamefont
  {L.}~\bibnamefont {Oroszl\'any}}, \bibinfo {author} {\bibfnamefont
  {A.}~\bibnamefont {Garc\'{\i}a-Fuente}}, \bibinfo {author} {\bibfnamefont
  {B.}~\bibnamefont {Ny\'ari}}, \bibinfo {author} {\bibfnamefont
  {L.}~\bibnamefont {Udvardi}}, \bibinfo {author} {\bibfnamefont
  {L.}~\bibnamefont {Szunyogh}}, \ and\ \bibinfo {author} {\bibfnamefont
  {J.}~\bibnamefont {Ferrer}},\ }\href {\doibase 10.1103/PhysRevB.108.214418}
  {\bibfield  {journal} {\bibinfo  {journal} {Phys. Rev. B}\ }\textbf {\bibinfo
  {volume} {108}},\ \bibinfo {pages} {214418} (\bibinfo {year}
  {2023}{\natexlab{b}})}\BibitemShut {NoStop}%
\bibitem [{\citenamefont {Sun}\ \emph {et~al.}(2023)\citenamefont {Sun},
  \citenamefont {Wang}, \citenamefont {Hu}, \citenamefont {Yang}, \citenamefont
  {Li}, \citenamefont {Huang}, \citenamefont {Li},\ and\ \citenamefont
  {Cheng}}]{sun2023theoretical}%
  \BibitemOpen
  \bibfield  {author} {\bibinfo {author} {\bibfnamefont {W.}~\bibnamefont
  {Sun}}, \bibinfo {author} {\bibfnamefont {W.}~\bibnamefont {Wang}}, \bibinfo
  {author} {\bibfnamefont {R.}~\bibnamefont {Hu}}, \bibinfo {author}
  {\bibfnamefont {C.}~\bibnamefont {Yang}}, \bibinfo {author} {\bibfnamefont
  {L.}~\bibnamefont {Li}}, \bibinfo {author} {\bibfnamefont {S.}~\bibnamefont
  {Huang}}, \bibinfo {author} {\bibfnamefont {X.}~\bibnamefont {Li}}, \ and\
  \bibinfo {author} {\bibfnamefont {Z.}~\bibnamefont {Cheng}},\ }\href
  {\doibase https://doi.org/10.1021/acsanm.3c03153} {\bibfield  {journal}
  {\bibinfo  {journal} {ACS Appl. Nano Mater.}\ }\textbf {\bibinfo {volume}
  {6}},\ \bibinfo {pages} {17021} (\bibinfo {year} {2023})}\BibitemShut
  {NoStop}%
\bibitem [{\citenamefont {Amoroso}\ \emph {et~al.}(2020)\citenamefont
  {Amoroso}, \citenamefont {Barone},\ and\ \citenamefont
  {Picozzi}}]{amoroso2020spontaneous}%
  \BibitemOpen
  \bibfield  {author} {\bibinfo {author} {\bibfnamefont {D.}~\bibnamefont
  {Amoroso}}, \bibinfo {author} {\bibfnamefont {P.}~\bibnamefont {Barone}}, \
  and\ \bibinfo {author} {\bibfnamefont {S.}~\bibnamefont {Picozzi}},\ }\href
  {\doibase https://doi.org/10.1038/s41467-020-19535-w} {\bibfield  {journal}
  {\bibinfo  {journal} {Nat. Commun.}\ }\textbf {\bibinfo {volume} {11}},\
  \bibinfo {pages} {5784} (\bibinfo {year} {2020})}\BibitemShut {NoStop}%
\bibitem [{\citenamefont {Poudel}\ \emph {et~al.}(2023)\citenamefont {Poudel},
  \citenamefont {Marmolejo-Tejada}, \citenamefont {Roll}, \citenamefont
  {Mosquera},\ and\ \citenamefont {Barraza-Lopez}}]{poudel2023creating}%
  \BibitemOpen
  \bibfield  {author} {\bibinfo {author} {\bibfnamefont {S.~P.}\ \bibnamefont
  {Poudel}}, \bibinfo {author} {\bibfnamefont {J.~M.}\ \bibnamefont
  {Marmolejo-Tejada}}, \bibinfo {author} {\bibfnamefont {J.~E.}\ \bibnamefont
  {Roll}}, \bibinfo {author} {\bibfnamefont {M.~A.}\ \bibnamefont {Mosquera}},
  \ and\ \bibinfo {author} {\bibfnamefont {S.}~\bibnamefont {Barraza-Lopez}},\
  }\href {\doibase 10.1103/PhysRevB.107.195128} {\bibfield  {journal} {\bibinfo
   {journal} {Phys. Rev. B}\ }\textbf {\bibinfo {volume} {107}},\ \bibinfo
  {pages} {195128} (\bibinfo {year} {2023})}\BibitemShut {NoStop}%
\bibitem [{\citenamefont {Wen}\ \emph {et~al.}(2020)\citenamefont {Wen},
  \citenamefont {Liu}, \citenamefont {Zhang}, \citenamefont {Xia},
  \citenamefont {Zhai}, \citenamefont {Zhang}, \citenamefont {Zhai},
  \citenamefont {Shen}, \citenamefont {He}, \citenamefont {Cheng} \emph
  {et~al.}}]{wen2020tunable}%
  \BibitemOpen
  \bibfield  {author} {\bibinfo {author} {\bibfnamefont {Y.}~\bibnamefont
  {Wen}}, \bibinfo {author} {\bibfnamefont {Z.}~\bibnamefont {Liu}}, \bibinfo
  {author} {\bibfnamefont {Y.}~\bibnamefont {Zhang}}, \bibinfo {author}
  {\bibfnamefont {C.}~\bibnamefont {Xia}}, \bibinfo {author} {\bibfnamefont
  {B.}~\bibnamefont {Zhai}}, \bibinfo {author} {\bibfnamefont {X.}~\bibnamefont
  {Zhang}}, \bibinfo {author} {\bibfnamefont {G.}~\bibnamefont {Zhai}},
  \bibinfo {author} {\bibfnamefont {C.}~\bibnamefont {Shen}}, \bibinfo {author}
  {\bibfnamefont {P.}~\bibnamefont {He}}, \bibinfo {author} {\bibfnamefont
  {R.}~\bibnamefont {Cheng}},  \emph {et~al.},\ }\href {\doibase
  https://dx.doi.org/10.1021/acs.nanolett.9b05128} {\bibfield  {journal}
  {\bibinfo  {journal} {Nano Lett.}\ }\textbf {\bibinfo {volume} {20}},\
  \bibinfo {pages} {3130} (\bibinfo {year} {2020})}\BibitemShut {NoStop}%
\bibitem [{\citenamefont {Fei}\ \emph {et~al.}(2018)\citenamefont {Fei},
  \citenamefont {Huang}, \citenamefont {Malinowski}, \citenamefont {Wang},
  \citenamefont {Song}, \citenamefont {Sanchez}, \citenamefont {Yao},
  \citenamefont {Xiao}, \citenamefont {Zhu}, \citenamefont {May} \emph
  {et~al.}}]{fei2018two}%
  \BibitemOpen
  \bibfield  {author} {\bibinfo {author} {\bibfnamefont {Z.}~\bibnamefont
  {Fei}}, \bibinfo {author} {\bibfnamefont {B.}~\bibnamefont {Huang}}, \bibinfo
  {author} {\bibfnamefont {P.}~\bibnamefont {Malinowski}}, \bibinfo {author}
  {\bibfnamefont {W.}~\bibnamefont {Wang}}, \bibinfo {author} {\bibfnamefont
  {T.}~\bibnamefont {Song}}, \bibinfo {author} {\bibfnamefont {J.}~\bibnamefont
  {Sanchez}}, \bibinfo {author} {\bibfnamefont {W.}~\bibnamefont {Yao}},
  \bibinfo {author} {\bibfnamefont {D.}~\bibnamefont {Xiao}}, \bibinfo {author}
  {\bibfnamefont {X.}~\bibnamefont {Zhu}}, \bibinfo {author} {\bibfnamefont
  {A.~F.}\ \bibnamefont {May}},  \emph {et~al.},\ }\href {\doibase
  https://doi.org/10.1038/s41563-018-0149-7} {\bibfield  {journal} {\bibinfo
  {journal} {Nat. Mater.}\ }\textbf {\bibinfo {volume} {17}},\ \bibinfo {pages}
  {778} (\bibinfo {year} {2018})}\BibitemShut {NoStop}%
\bibitem [{\citenamefont {Keeney}\ \emph {et~al.}(2013)\citenamefont {Keeney},
  \citenamefont {Maity}, \citenamefont {Schmidt}, \citenamefont {Amann},
  \citenamefont {Deepak}, \citenamefont {Petkov}, \citenamefont {Roy},
  \citenamefont {Pemble},\ and\ \citenamefont {Whatmore}}]{keeney2013magnetic}%
  \BibitemOpen
  \bibfield  {author} {\bibinfo {author} {\bibfnamefont {L.}~\bibnamefont
  {Keeney}}, \bibinfo {author} {\bibfnamefont {T.}~\bibnamefont {Maity}},
  \bibinfo {author} {\bibfnamefont {M.}~\bibnamefont {Schmidt}}, \bibinfo
  {author} {\bibfnamefont {A.}~\bibnamefont {Amann}}, \bibinfo {author}
  {\bibfnamefont {N.}~\bibnamefont {Deepak}}, \bibinfo {author} {\bibfnamefont
  {N.}~\bibnamefont {Petkov}}, \bibinfo {author} {\bibfnamefont
  {S.}~\bibnamefont {Roy}}, \bibinfo {author} {\bibfnamefont {M.~E.}\
  \bibnamefont {Pemble}}, \ and\ \bibinfo {author} {\bibfnamefont {R.~W.}\
  \bibnamefont {Whatmore}},\ }\href {\doibase
  https://doi.org/10.1111/jace.12467} {\bibfield  {journal} {\bibinfo
  {journal} {J. Am. Ceram. Soc.}\ }\textbf {\bibinfo {volume} {96}},\ \bibinfo
  {pages} {2339} (\bibinfo {year} {2013})}\BibitemShut {NoStop}%
\bibitem [{\citenamefont {Soler}\ \emph {et~al.}(2002)\citenamefont {Soler},
  \citenamefont {Artacho}, \citenamefont {Gale}, \citenamefont {Garc{\'\i}a},
  \citenamefont {Junquera}, \citenamefont {Ordej{\'o}n},\ and\ \citenamefont
  {S{\'a}nchez-Portal}}]{siesta}%
  \BibitemOpen
  \bibfield  {author} {\bibinfo {author} {\bibfnamefont {J.~M.}\ \bibnamefont
  {Soler}}, \bibinfo {author} {\bibfnamefont {E.}~\bibnamefont {Artacho}},
  \bibinfo {author} {\bibfnamefont {J.~D.}\ \bibnamefont {Gale}}, \bibinfo
  {author} {\bibfnamefont {A.}~\bibnamefont {Garc{\'\i}a}}, \bibinfo {author}
  {\bibfnamefont {J.}~\bibnamefont {Junquera}}, \bibinfo {author}
  {\bibfnamefont {P.}~\bibnamefont {Ordej{\'o}n}}, \ and\ \bibinfo {author}
  {\bibfnamefont {D.}~\bibnamefont {S{\'a}nchez-Portal}},\ }\href {\doibase
  https://doi.org/10.1088/0953-8984/14/11/302} {\bibfield  {journal} {\bibinfo
  {journal} {J. Phys. Condens. Matter}\ }\textbf {\bibinfo {volume} {14}},\
  \bibinfo {pages} {2745} (\bibinfo {year} {2002})}\BibitemShut {NoStop}%
\bibitem [{\citenamefont {Hamann}(2013)}]{norm_conserving}%
  \BibitemOpen
  \bibfield  {author} {\bibinfo {author} {\bibfnamefont {D.}~\bibnamefont
  {Hamann}},\ }\href {\doibase https://doi.org/10.1103/PhysRevB.88.085117}
  {\bibfield  {journal} {\bibinfo  {journal} {Phys. Rev. B}\ }\textbf {\bibinfo
  {volume} {88}},\ \bibinfo {pages} {085117} (\bibinfo {year}
  {2013})}\BibitemShut {NoStop}%
\bibitem [{\citenamefont {Garc{\'\i}a}\ \emph {et~al.}(2018)\citenamefont
  {Garc{\'\i}a}, \citenamefont {Verstraete}, \citenamefont {Pouillon},\ and\
  \citenamefont {Junquera}}]{psml}%
  \BibitemOpen
  \bibfield  {author} {\bibinfo {author} {\bibfnamefont {A.}~\bibnamefont
  {Garc{\'\i}a}}, \bibinfo {author} {\bibfnamefont {M.~J.}\ \bibnamefont
  {Verstraete}}, \bibinfo {author} {\bibfnamefont {Y.}~\bibnamefont
  {Pouillon}}, \ and\ \bibinfo {author} {\bibfnamefont {J.}~\bibnamefont
  {Junquera}},\ }\href {\doibase https://doi.org/10.1016/j.cpc.2018.02.011}
  {\bibfield  {journal} {\bibinfo  {journal} {Comput. Phys. Commun.}\ }\textbf
  {\bibinfo {volume} {227}},\ \bibinfo {pages} {51} (\bibinfo {year}
  {2018})}\BibitemShut {NoStop}%
\bibitem [{\citenamefont {Van~Setten}\ \emph {et~al.}(2018)\citenamefont
  {Van~Setten}, \citenamefont {Giantomassi}, \citenamefont {Bousquet},
  \citenamefont {Verstraete}, \citenamefont {Hamann}, \citenamefont {Gonze},\
  and\ \citenamefont {Rignanese}}]{pseudodojo}%
  \BibitemOpen
  \bibfield  {author} {\bibinfo {author} {\bibfnamefont {M.}~\bibnamefont
  {Van~Setten}}, \bibinfo {author} {\bibfnamefont {M.}~\bibnamefont
  {Giantomassi}}, \bibinfo {author} {\bibfnamefont {E.}~\bibnamefont
  {Bousquet}}, \bibinfo {author} {\bibfnamefont {M.~J.}\ \bibnamefont
  {Verstraete}}, \bibinfo {author} {\bibfnamefont {D.~R.}\ \bibnamefont
  {Hamann}}, \bibinfo {author} {\bibfnamefont {X.}~\bibnamefont {Gonze}}, \
  and\ \bibinfo {author} {\bibfnamefont {G.-M.}\ \bibnamefont {Rignanese}},\
  }\href {\doibase https://doi.org/10.1016/j.cpc.2018.01.012} {\bibfield
  {journal} {\bibinfo  {journal} {Comput. Phys. Commun.}\ }\textbf {\bibinfo
  {volume} {226}},\ \bibinfo {pages} {39} (\bibinfo {year} {2018})}\BibitemShut
  {NoStop}%
\bibitem [{\citenamefont {Papior}\ \emph {et~al.}(2018)\citenamefont {Papior},
  \citenamefont {Calogero},\ and\ \citenamefont
  {Brandbyge}}]{papior2018simple}%
  \BibitemOpen
  \bibfield  {author} {\bibinfo {author} {\bibfnamefont {N.~R.}\ \bibnamefont
  {Papior}}, \bibinfo {author} {\bibfnamefont {G.}~\bibnamefont {Calogero}}, \
  and\ \bibinfo {author} {\bibfnamefont {M.}~\bibnamefont {Brandbyge}},\ }\href
  {\doibase https://doi.org/10.1088/1361-648X/aac4dd} {\bibfield  {journal}
  {\bibinfo  {journal} {J. Phys.: Condens. Matter}\ }\textbf {\bibinfo {volume}
  {30}},\ \bibinfo {pages} {25LT01} (\bibinfo {year} {2018})}\BibitemShut
  {NoStop}%
\bibitem [{\citenamefont {Gonze}\ \emph {et~al.}(2009)\citenamefont {Gonze},
  \citenamefont {Amadon}, \citenamefont {Anglade}, \citenamefont {Beuken},
  \citenamefont {Bottin}, \citenamefont {Boulanger}, \citenamefont {Bruneval},
  \citenamefont {Caliste}, \citenamefont {Caracas}, \citenamefont
  {C{\^o}t{\'e}} \emph {et~al.}}]{gonze2009abinit}%
  \BibitemOpen
  \bibfield  {author} {\bibinfo {author} {\bibfnamefont {X.}~\bibnamefont
  {Gonze}}, \bibinfo {author} {\bibfnamefont {B.}~\bibnamefont {Amadon}},
  \bibinfo {author} {\bibfnamefont {P.-M.}\ \bibnamefont {Anglade}}, \bibinfo
  {author} {\bibfnamefont {J.-M.}\ \bibnamefont {Beuken}}, \bibinfo {author}
  {\bibfnamefont {F.}~\bibnamefont {Bottin}}, \bibinfo {author} {\bibfnamefont
  {P.}~\bibnamefont {Boulanger}}, \bibinfo {author} {\bibfnamefont
  {F.}~\bibnamefont {Bruneval}}, \bibinfo {author} {\bibfnamefont
  {D.}~\bibnamefont {Caliste}}, \bibinfo {author} {\bibfnamefont
  {R.}~\bibnamefont {Caracas}}, \bibinfo {author} {\bibfnamefont
  {M.}~\bibnamefont {C{\^o}t{\'e}}},  \emph {et~al.},\ }\href {\doibase
  https://doi.org/10.1016/j.cpc.2009.07.007} {\bibfield  {journal} {\bibinfo
  {journal} {Comput. Phys. Commun.}\ }\textbf {\bibinfo {volume} {180}},\
  \bibinfo {pages} {2582} (\bibinfo {year} {2009})}\BibitemShut {NoStop}%
\bibitem [{\citenamefont {Monkhorst}\ and\ \citenamefont {Pack}(1976)}]{mp}%
  \BibitemOpen
  \bibfield  {author} {\bibinfo {author} {\bibfnamefont {H.~J.}\ \bibnamefont
  {Monkhorst}}\ and\ \bibinfo {author} {\bibfnamefont {J.~D.}\ \bibnamefont
  {Pack}},\ }\href {\doibase https://doi.org/10.1103/PhysRevB.13.5188}
  {\bibfield  {journal} {\bibinfo  {journal} {Phys. Rev. B}\ }\textbf {\bibinfo
  {volume} {13}},\ \bibinfo {pages} {5188} (\bibinfo {year}
  {1976})}\BibitemShut {NoStop}%
\bibitem [{\citenamefont {Perdew}\ \emph {et~al.}(1996)\citenamefont {Perdew},
  \citenamefont {Burke},\ and\ \citenamefont {Ernzerhof}}]{pbe}%
  \BibitemOpen
  \bibfield  {author} {\bibinfo {author} {\bibfnamefont {J.~P.}\ \bibnamefont
  {Perdew}}, \bibinfo {author} {\bibfnamefont {K.}~\bibnamefont {Burke}}, \
  and\ \bibinfo {author} {\bibfnamefont {M.}~\bibnamefont {Ernzerhof}},\ }\href
  {\doibase 10.1103/PhysRevLett.77.3865} {\bibfield  {journal} {\bibinfo
  {journal} {Phys. Rev. Lett.}\ }\textbf {\bibinfo {volume} {77}},\ \bibinfo
  {pages} {3865} (\bibinfo {year} {1996})}\BibitemShut {NoStop}%
\bibitem [{\citenamefont {Grimme}\ \emph {et~al.}(2010)\citenamefont {Grimme},
  \citenamefont {Antony}, \citenamefont {Ehrlich},\ and\ \citenamefont
  {Krieg}}]{grimme2010consistent}%
  \BibitemOpen
  \bibfield  {author} {\bibinfo {author} {\bibfnamefont {S.}~\bibnamefont
  {Grimme}}, \bibinfo {author} {\bibfnamefont {J.}~\bibnamefont {Antony}},
  \bibinfo {author} {\bibfnamefont {S.}~\bibnamefont {Ehrlich}}, \ and\
  \bibinfo {author} {\bibfnamefont {H.}~\bibnamefont {Krieg}},\ }\href
  {https://doi.org/10.1063/1.3382344} {\bibfield  {journal} {\bibinfo
  {journal} {J. Chem. Phys.}\ }\textbf {\bibinfo {volume} {132}} (\bibinfo
  {year} {2010})}\BibitemShut {NoStop}%
\bibitem [{\citenamefont {Neugebauer}\ and\ \citenamefont
  {Scheffler}(1992)}]{dipole_correction_1}%
  \BibitemOpen
  \bibfield  {author} {\bibinfo {author} {\bibfnamefont {J.}~\bibnamefont
  {Neugebauer}}\ and\ \bibinfo {author} {\bibfnamefont {M.}~\bibnamefont
  {Scheffler}},\ }\href {\doibase https://doi.org/10.1103/PhysRevB.46.16067}
  {\bibfield  {journal} {\bibinfo  {journal} {Phys. Rev. B}\ }\textbf {\bibinfo
  {volume} {46}},\ \bibinfo {pages} {16067} (\bibinfo {year}
  {1992})}\BibitemShut {NoStop}%
\bibitem [{\citenamefont {Rutter}(2018)}]{rutter2018c2x}%
  \BibitemOpen
  \bibfield  {author} {\bibinfo {author} {\bibfnamefont {M.~J.}\ \bibnamefont
  {Rutter}},\ }\href {\doibase doi={https://doi.org/10.1016/j.cpc.2017.12.008}}
  {\bibfield  {journal} {\bibinfo  {journal} {Comput. Phys. Commun.}\ }\textbf
  {\bibinfo {volume} {225}},\ \bibinfo {pages} {174} (\bibinfo {year}
  {2018})}\BibitemShut {NoStop}%
\end{thebibliography}

%

\clearpage

\includepdf[pages={1}]{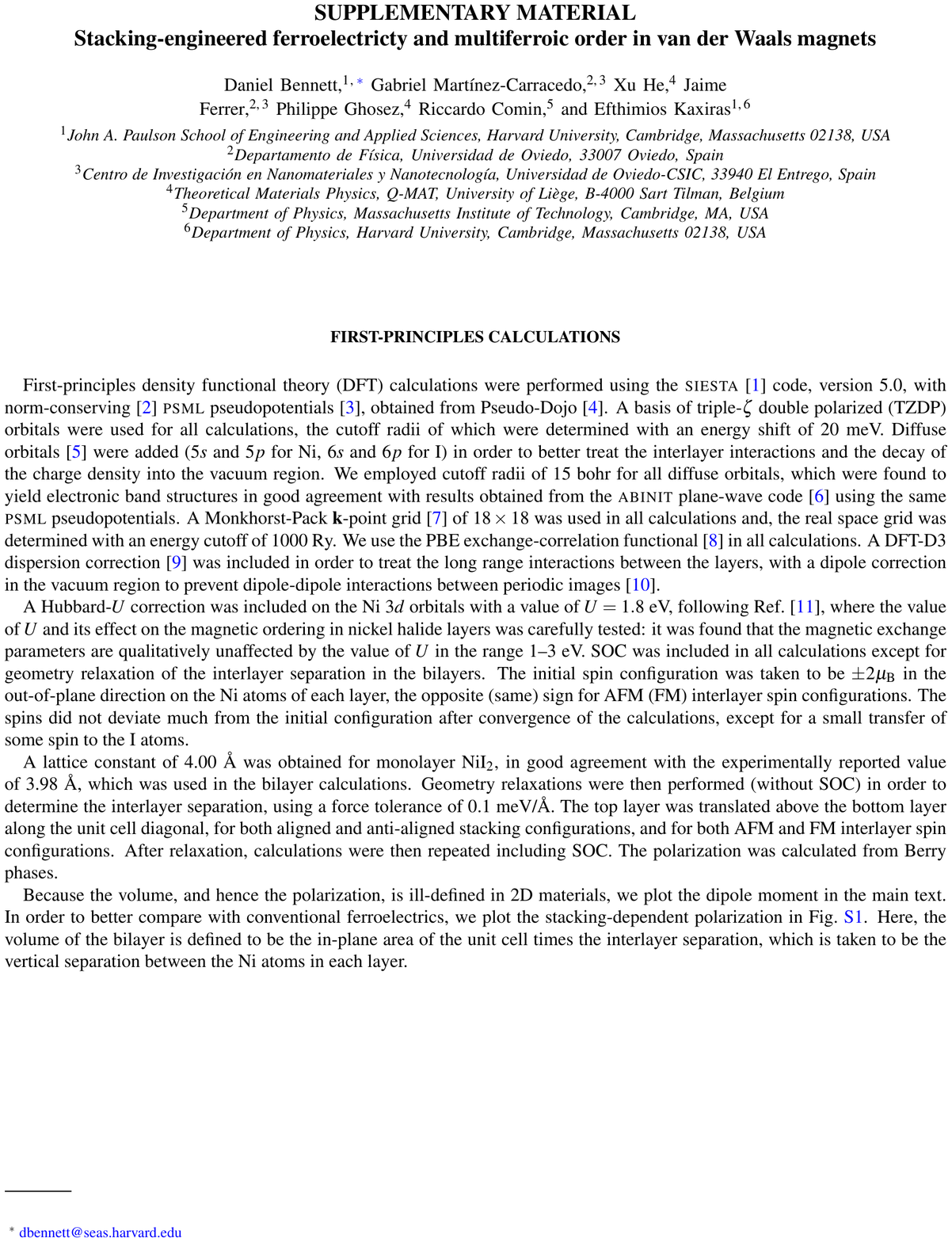}
\clearpage
\includepdf[pages={2}]{./SM.pdf}
\clearpage
\includepdf[pages={3}]{./SM.pdf}
\clearpage
\includepdf[pages={4}]{./SM.pdf}
\clearpage
\includepdf[pages={5}]{./SM.pdf}
\clearpage
\includepdf[pages={6}]{./SM.pdf}
\clearpage
\includepdf[pages={7}]{./SM.pdf}
\clearpage
\includepdf[pages={8}]{./SM.pdf}
\clearpage
\includepdf[pages={9}]{./SM.pdf}
\clearpage
\includepdf[pages={10}]{./SM.pdf}
\clearpage
\includepdf[pages={11}]{./SM.pdf}
\clearpage
\includepdf[pages={12}]{./SM.pdf}
\clearpage
\includepdf[pages={13}]{./SM.pdf}
\clearpage
\includepdf[pages={14}]{./SM.pdf}
\clearpage
\includepdf[pages={15}]{./SM.pdf}

\end{document}